\documentclass[a4paper,twocolumn,prb,superscriptaddress,showpacs]{revtex4}

\usepackage{dcolumn}
\usepackage{amsmath}
\usepackage{graphicx}
\usepackage{latexsym}  
\usepackage{amsfonts}  
\usepackage{amssymb}
\usepackage{color}  
\usepackage{bbm}
\DeclareGraphicsExtensions{.eps,.pdf,.png,.gif,.jpg}  
\usepackage{comment}

\newcommand{\bo}{\mathbf}
\newcommand{\te}{\textrm}
\newcommand{\be}{\begin{equation}}
\newcommand{\ee}{\end{equation}}
\newcommand{\bea}{\begin{eqnarray}}
\newcommand{\eea}{\end{eqnarray}}
\newcommand{\w}{\omega}
\newcommand{\g}{\gamma}

\newcommand{\vg} {{\bf g}}
\newcommand{\vn} {{\bf n}}
\newcommand{\vG} {{\bf G}}
\newcommand{\vH} {{\bf H}}
\newcommand{\vSi} {{\bf \Sigma}}
\newcommand{\nn} {\nonumber}

\def\a{\alpha}
\def\s{\sigma}

\begin{document}

\title{Correlation effects in bistability at the nanoscale: steady state and 
beyond}

\author{E. Khosravi}
\affiliation{Max-Planck Institut f\"ur Mikrostrukturphysik, Weinberg 2, 
D-06120 Halle, Germany}
\affiliation{European Theoretical Spectroscopy Facility (ETSF)}

\author{A.-M. Uimonen}
\affiliation{ Department of Physics, Nanoscience Center, FIN 40014, 
University of Jyv\"askyl\"a, Jyv\"askyl\"a, Finland}
\affiliation{European Theoretical Spectroscopy Facility (ETSF)}

\author{A. Stan}
\affiliation{ Department of Physics, Nanoscience Center, FIN 40014, 
University of Jyv\"askyl\"a, Jyv\"askyl\"a, Finland}
\affiliation{European Theoretical Spectroscopy Facility (ETSF)}

\author{G. Stefanucci}
\affiliation{Dipartimento di Fisica, Universit\`{a} di Roma Tor Vergata,
Via della Ricerca Scientifica 1, 00133 Rome, Italy}
\affiliation{INFN, Laboratori Nazionali di Frascati, Via E. Fermi 40, 00044 Frascati, Italy}
\affiliation{European Theoretical Spectroscopy Facility (ETSF)}
 
\author{S. Kurth}  
\affiliation{Nano-Bio Spectroscopy Group, 
Departamento de F\'{i}sica de Materiales, 
Universidad del Pa\'{i}s Vasco UPV/EHU, Centro F\'{i}sica de Materiales 
CSIC-UPV/EHU, Avenida de Tolosa 72, E-20018 San Sebasti\'{a}n, Spain} 
\affiliation{IKERBASQUE, Basque Foundation for Science, E-48011 Bilbao, Spain}
\affiliation{European Theoretical Spectroscopy Facility (ETSF)}
  
\author{R. van Leeuwen}
\affiliation{ Department of Physics, Nanoscience Center, FIN 40014, 
University of Jyv\"askyl\"a, Jyv\"askyl\"a, Finland}
\affiliation{European Theoretical Spectroscopy Facility (ETSF)}

\author{E. K. U. Gross}
\affiliation{Max-Planck Institut f\"ur Mikrostrukturphysik, Weinberg 2, 
D-06120 Halle, Germany}
\affiliation{European Theoretical Spectroscopy Facility (ETSF)}

\date{\today}  

\begin{abstract}
The possibility of finding multistability in the density and current 
of an interacting nanoscale junction 
coupled to semi-infinite leads is studied at various levels of approximation. 
The system is driven out of equilibrium by an external bias and  
the non-equilibrium properties  are determined by real-time 
propagation  using  both time-dependent density functional theory 
(TDDFT) and many-body perturbation theory (MBPT).
In TDDFT the exchange-correlation effects are described within a 
recently proposed adiabatic local density approximation (ALDA). In MBPT the 
electron-electron interaction is incorporated in a many-body 
self-energy which is then approximated at the Hartree-Fock (HF), second-Born (2B)
and $GW$ level. Assuming the existence of a steady-state and 
solving directly the steady-state equations 
we find multiple solutions in the HF 
approximation and within the ALDA. In these cases we investigate if 
and how these solutions can be reached through time evolution and how 
to reversibly switch between them. 
We further show that for the same cases the inclusion of dynamical correlation effects suppresses 
bistability.

\end{abstract}

  
\maketitle  

\section{Introduction}
\label{sec:intro}
The phenomenon of bistability has been the subject of 
several studies in the field of quantum transport.
In their seminal paper Goldman {\it et al.}~\cite{Goldman_PRL_87} reported 
the observation of bistability in the I-V curve of double-barrier resonant 
tunneling (DBRT) structures, thus stimulating many theoretical 
\cite{sf.1990,fj.1992,pf.1993,DaiNi05} and experimental investigations 
\cite{sheard__APL_88,ZGTC:88} on the subject. The  bistability  is 
a non-linear effect induced by the electrostatic charge build-up in 
the quantum well  and  occurs in the bias window of negative 
differential resistance.\cite{Goldman_PRL_87}
From the theoretical point of view, various techniques have been used
to capture this phenomenon, ranging from a crude estimate of the 
charge build-up \cite{sheard__APL_88} to self-consistent calculations at a 
mean field-level. 
\cite{sf.1990,fj.1992,pf.1993,zwlhcg.1994,Zhang:01,DzhioevKosov:11}

With the increasing interest in transport through nanoscale devices, 
in particular using molecules as a possible 
component of future electronic circuits, 
the study of intrinsic bistability in nanoelectronics  has gained new attention.
The possibility of finding molecular devices equivalent to conventional non-linear devices, 
 such as diodes and transistors, is indeed an attractive perspective.
There have already been some successful attempts along these lines. 
Molecular devices with large on-off current ratio and large negative differential resistance,
 behaving similarly to mesoscopic DBRT structures, have been reported. 
\cite{ChenReedRawlettTout:99,ChenReed:02}
So far, the great majority of bistability studies have been limited 
to the steady-state regime and  performed  
within the framework of the Landauer formalism combined with static density 
functional theory (DFT). At the Hartree level, bistability was reported  
for a double quantum dot structure.\cite{Negre_CPL_08,SSSBHT:06}
In the context of time-dependent (TD) DFT, the inclusion of memory effects 
beyond the adiabatic approximation 
is not straightforward and the development of accurate functionals to 
be used in numerical calculations is still under way. 
A promising and timely, even though computationally demanding, alternative is 
the solution of the Kadanoff-Baym (KB) equations
\cite{KadanoffBaym:62,dvl.2007,sdvl.2009,MyohanenStanStefanucciLeeuwen:08, MyohanenStanStefanucciLeeuwen:09,dahlen06procNGFT,dahlen06procKB}  
using self-energies from many-body perturbation theory (MBPT).
The advantage of MBPT  over TDDFT is the inclusion of  dynamical 
exchange-correlation (XC) effects, i.e., effects arising from a 
frequency-dependent self-energy, 
in a more systematic way through the selection of suitable Feynman diagrams.
Thus, MBPT provides an important 
tool to go beyond the commonly used adiabatic approximations and to quantify 
the importance of dynamical XC effects.

The fundamental issue which we address in this paper is 
whether the bistability phenomenon found in static DFT, Hartree and 
Hartree-Fock (HF) approximation survives when  dynamical XC 
effects  are taken into account. 
In contrast to DFT and mean-field approximations the steady-state 
equations of MBPT do not form a 
closed set of equations for the density only. This difficulty renders 
the search for the bistability regime in MBPT computationally very costly.
To overcome the problem we implement a time-dependent strategy.\cite{MyohanenStanStefanucciLeeuwen:08,MyohanenStanStefanucciLeeuwen:09,petri:10a,MBTDDFT:11,
BaerSeidemanIlaniNeuhauser:04,StefanucciAlmbladh:04,StefanucciAlmbladh:04-2,ksarg.2005,BurkeCarGebauer:05,Zheng:10,Yametal:2011,zheng:09,zheng:08,jin:08}
We first solve the steady-state equations of DFT and mean-field 
theory to determine the parameter range for bistability. Then we 
go beyond the current state-of-the-art and provide a TD
description of the bistability phenomenon in adiabatic TDDFT
\cite{BaerSeidemanIlaniNeuhauser:04,StefanucciAlmbladh:04,
StefanucciAlmbladh:04-2,ksarg.2005,BurkeCarGebauer:05,Zheng:10,Yametal:2011} and 
TD mean field theory. We show how  to switch 
between different stable states by means of ultrafast  
gate voltages or external biases (driving fields). The possibility of 
reversibly switching between different stable steady-states is an 
aspect that has remained largely unexplored.\cite{proceedings}
Knowing how to steer the electron dynamics in real time 
we use the same driving fields in correlated MBPT 
simulations. The calculations are performed with the fully self-consistent second-Born (2B) and 
$GW$ approximations which have recently been shown\cite{{MBTDDFT:11}} to agree with 
numerically exact TD-DMRG 
results.\cite{HeidrichMeisnerFeiguinDagotto:09} 
In all cases studied here 
where adiabatic DFT and HF theory predict bistability 
{\em dynamical XC effects destroy the phenomenon}. 

The paper is organized as follows:
In Section \ref{model}  we introduce the model used in our study. 
In Section \ref{tddft} we introduce the TDDFT approach to transport, with a 
particular emphasis on the real-time propagation to study time-dependent 
transport phenomena. Assuming that the system reaches a steady 
state, the steady-state density can be calculated 
without explicitly propagating in time, as explained in 
Section \ref{steady-state}. In this Section a separate 
(fixed-point) analysis is given to determine whether or not a solution is 
stable. 
As an alternative real-time approach to transport, we introduce in 
Section \ref{mbpt} the MBPT approach based on the solution of the 
KB equations. 
In Section \ref{result}, we present  the results of our numerical 
simulations 
for a certain set of parameters for which multiple solutions are 
observed in DFT.
Finally,  conclusions are drawn in Section \ref{conclus}.

\section{The model}
\label{model}

In this work we study multistability in quantum transport, i.e., we study the 
question if, for a given set of parameters, a biased system can have more than 
one steady state. Apart from proposing a way to classify the different steady 
states as stable or unstable, we also investigate by explicit time 
evolution the possibility to reversibly switch between different steady states 
by application of a suitably chosen, time-dependent perturbation. 

These questions are addressed in model systems. We consider a nanoscale 
device consisting of a few interacting sites contacted to two non-interacting 
tight-binding leads. Initially, the contacted system is in equilibrium at 
a given temperature and chemical potential. At time $t_0=0$ we switch on a 
bias in the leads and follow the time evolution of the perturbed system.

The Hamiltonian of the system consists of three different parts 
\be
\hat{H}(t) = \hat{H}_{C}(t) + \sum_{\alpha=L,R} \hat{H}_{\alpha}(t) + 
\hat{H}_{T},
\label{hamil}
\ee
where the Hamiltonian $\hat{H}_{C}(t)$ of the central device describes a chain
of $N_C$ sites with a Hubbard-type on-site electron-electron interaction:
\bea
\hat{H}_{C}(t) &=& \sum_{\substack{ i=1 \\\s}}^{N_C}  \varepsilon^{C}_{i}(t) 
\hat{d}_{i\sigma}^{\dagger} \hat{d}_{i\sigma} - 
\sum_{\substack{ i=1 \\\s}}^{N_C-1} ( V_{C} \hat{d}_{i\sigma}^{\dagger} 
\hat{d}_{i+1\sigma} + H.c.) \nn \\
&& + \frac{1}{2} \sum_{\substack{ i=1 \\\s\s'}}^{N_C} 
U\hat{d}_{i\sigma}^{\dagger} \hat{d}_{i\sigma'}^{\dagger} 
\hat{d}_{i\sigma'} \hat{d}_{i\sigma} \; .
\label{hamilc}
\eea
Here, $\hat{d}_{i\sigma}^{\dagger}$ ($\hat{d}_{i\sigma}$) denote creation 
(annihilation) operators for electrons with spin $\sigma$ at site $i$. The 
$\varepsilon^{C}_{i}(t)$ are on-site energies which may consist 
of an arbitrary time-dependent part, denoted as $V_{g,i}(t)$, plus a time-independent part,~$\varepsilon^{C}_{i}$. The nearest-neighbor hopping 
in the chain is $V_{C}$ and $U$ is the on-site Hubbard interaction. 

The non-interacting left ($L$) and right ($R$) leads, $\alpha=L,R$, are 
described by one-dimensional semi-infinite chains with Hamiltonian
\bea
\hat{H}_{\alpha}(t) &=& \sum_{\substack{ i=1 \\\s}}^{\infty} \left( 
\varepsilon_{\alpha} + W_{\alpha}(t) \right) 
\hat{c}_{i\sigma\alpha}^{\dagger} \hat{c}_{i\sigma\alpha} \nn \\
&& -  \sum_{\substack{ i=1 \\\s}}^{\infty} \left( V_{\alpha} 
\hat{c}_{i\sigma\alpha}^{\dagger} \hat{c}_{i+1\sigma\alpha} + H.c. \right), 
\eea
with creation (annihilation) operators $\hat{c}_{i\sigma\alpha}^{\dagger} $ 
($\hat{c}_{i\sigma\alpha}$) for electrons with spin $\sigma$ at site $i$ in 
the lead $\alpha$. The on-site energies $\varepsilon_{\alpha}$ and the hopping 
matrix elements $V_{\alpha}$ are independent of time and site index while 
$W_{\alpha}(t)$ describes a time-dependent, site-independent bias applied to 
the lead $\alpha$. 

Finally, the tunneling Hamiltonian which couples the leads to the device 
is given by 
\be
\hat{H}_{T} = - \sum_{\sigma} \left( 
V_{\rm link} \hat{d}_{1\sigma}^{\dagger} \hat{c}_{1\sigma L} 
+ V_{\rm link} \hat{d}_{N_C \sigma}^{\dagger} \hat{c}_{1\sigma R}  
+ H.c. \right),
\ee
where we have adopted the convention that the site in the lead $\alpha$ 
connected to the device is labeled by the site index 1 and $V_{\rm link}$ is 
the hopping between the central region and the leads. 

\section{Multistability in time-dependent density functional theory}
\label{tddft}

In this section we introduce the TDDFT approach~\cite{TDDFT_book} which we 
will use to investigate multistability in our model system. 
In TDDFT, the complicated problem of interacting electrons is mapped, in 
principle exactly, on the much simpler problem of  
non-interacting Kohn-Sham (KS) electrons moving in an effective local 
potential, thus providing a natural way  
to account for correlation effects in both leads and device 
region.\cite{StefanucciAlmbladh:04,StefanucciAlmbladh:04-2} Since the 
method only involves 
the propagation of  single-particle wavefunctions, it 
promises for a computationally efficient way to study time-dependent phenomena 
in quantum transport.\cite{ksarg.2005} The real-time TDDFT approach to 
transport will be described in Sec.~\ref{tddftsec}.
While in principle exact, in practice TDDFT 
requires the use of approximations for the time-dependent XC functionals. The most popular one, the adiabatic 
local density approximation (ALDA), depends only on the {\em local} and 
{\em instantaneous} density, i.e., it does not include memory effects. 
On one hand this feature is certainly a shortcoming of the approximation  
whose consequences for time-dependent transport still need to be explored. 
On the other hand, in the context of multistability, this approximation 
allows one to formulate the steady-state condition in terms of a  
closed set of non-linear equations for the steady-state density. The solution of these equations allows for an 
efficient scan of parameter space to find those parameter values for which 
multistability is possible. This steady-state approach with the local and 
adiabatic approximation of TDDFT will be described in Sec.~\ref{steady-state}. 

\subsection{Real-time TDDFT for transport}
\label{tddftsec}

One of the technical difficulties in applying TDDFT to quantum transport 
lies in the necessity of propagating an infinite non-periodic system, or 
equivalently, a finite open system attached to semi-infinite leads. Here we 
sketch the technique how this can be achieved. The technical details of the 
algorithm can be found in Ref.~\onlinecite{ksarg.2005}. 

In a localized (site) basis, the KS Hamiltonian of the total system consisting 
of left lead, device, and right lead can be partitioned in a block-diagonal 
matrix form as
\be
\left(
\begin{array}{ccc} 
\vH^{KS}_{LL}(t) & \vH_{LC} & 0 \\
\vH_{CL} & \vH^{KS}_{CC}(t) & \vH_{CR} \\
0 & \vH_{RC} & \vH^{KS}_{RR}(t) \\
\end{array} \right),
\label{hamil_matrix}
\ee
where $\vH^{KS}_{CC}(t)$  is the Hamiltonian of the isolated device and  
$\vH^{KS}_{\a \a}(t)=\vH^{KS}_{\a \a}+\bo{W}^{KS}_{\a}(t)$ is the Hamiltonian of 
the isolated lead $\alpha$. The time-dependent potential in the leads has the 
simple form $\bo{W}^{KS}_{\a}(t)=W_{\a} (t) \mathbbm{1}_{\a} $ where 
$\mathbbm{1}_{\a}$ is the identity matrix for lead $\a = L,R$. 
Finally, $\vH_{C \alpha}$ and $\vH_{\alpha C}$ describe the coupling between lead 
$\alpha$ and the device. Here and in the following we use boldface 
notation to indicate matrices in one-electron labels. 

Using downfolding techniques one can derive the equation of motion for the 
$k$-th KS single-particle orbital projected onto the central region, 
$\psi_{k,C}(t)$, which reads
\bea
\left[ i \partial_t - \vH_{CC}^{KS}(t) \right] \psi_{k,C}(t) = 
\int_0^t {\rm d} \bar{t} \; \vSi_{\rm em}^{R}(t,\bar{t}) \psi_{k,C}(\bar{t}) 
\nn\\
+ \sum_{\alpha} \vH_{C \alpha} \vg_{\alpha \alpha}^R(t,0) \psi_{k,\alpha}(0) \; . 
\label{eq:eqmTDDFT}
\eea
Here, $\psi_{k,\alpha}(0)$ is the projection of the $k$-th KS orbital on lead 
$\alpha$ at the initial time $t_0=0$, $\vg_{\alpha \alpha}^{R}(t,t')$ is the 
retarded Green function of the isolated lead $\alpha$ and the retarded embedding 
self-energy is defined as 
\be
\vSi^{R}_{\rm em}(t,t') = \sum_{\alpha=L,R} \vH_{C \alpha} \vg^{R}_{\alpha 
\alpha}(t,t') \vH_{\alpha C} \; .
\label{sigma_emb}
\ee
The  time-dependent density at site $j$ of region $C$ at zero temperature can 
be written as  
\be
n_{j}(t) = 2 \sum_{k}^{\rm occ} | \psi_{k,C}(j,t)|^2,
\label{dens_c}
\ee
where the sum runs over the occupied KS orbitals 
and the prefactor is due to spin degeneracy. 

In our model the KS Hamiltonian matrix $\vH_{CC}^{KS}(t)$ has a tridiagonal 
form where the only non-vanishing entries are the off-diagonal matrix elements 
$[\vH_{CC}^{KS}(t)]_{j,j+1} = [\vH_{CC}^{KS}(t)]_{j+1,j} = -V_{C}$ with 
$j=1,\cdots,N_C-1$, and the diagonal matrix elements
\bea
[\vH_{CC}^{KS}(t)]_{jj} &=& v_{\rm KS}(j,t) \nn \\ 
&=& \varepsilon_j^C(t) + v_H[n_j(t)]+ v_{\rm xc}[n](j,t) ,
\label{kspot}
\eea
with $j=1,\cdots,N_C$. The second term on the r.h.s of 
Eq.~(\ref{kspot}) is the Hartree potential and the third term is the 
XC potential of TDDFT for model systems.\cite{spc.2010,t.2011} Although in our model the 
interaction is restricted to the device region only, the exact XC potential 
has contributions in the adjacent lead regions as well and will rigorously 
vanish only deep inside the leads.\cite{evers:2004,Schenk:11} Therefore, already at this point we make 
an approximation by restricting the XC potential to the device region only.

Of course, the exact form of $v_{\rm xc}[n]$ is unknown and in practice one has 
to resort to approximations. For lattice systems, a local density 
approximation (LDA) based on the Bethe ansatz solution of the uniform 
one-dimensional Hubbard model has been suggested 
\cite{SchoenhammerGunnarssonNoack:95,LimaSilvaOliveiraCapelle:03} and 
a parameterization of this Bethe ansatz LDA (BALDA) has been presented in 
Ref.~\onlinecite{LimaSilvaOliveiraCapelle:03}. The adiabatic 
version\cite{Verdozzi:08} of this functional (ABALDA) makes $v_{\rm xc}[n]$ local in 
both space and time, i.e., $v_{\rm xc}[n](j,t) = v_{\rm xc}(n_{j}(t))$. 
The original BALDA was designed for the uniform Hubbard model with Hubbard 
interaction $U$ and nearest-neighbor hopping $V$ everywhere. 
A modification of this functional for the case of a single interacting 
impurity site ($N_C=1$) connected by a hopping matrix element $V_{\rm link}$ to 
the leads with hopping $V$ has been suggested in 
Ref.~\onlinecite{KurthStefanucciKhosraviVerdozziGross:10}.
In this work we use this modified functional both for the case of a single 
impurity \cite{MBTDDFT:11} and also for $N_C>1$ where in the latter case we 
impose the restriction that the hopping $V_{C}$ between the sites in the 
central region is equal to the hopping $V_{\textrm{link}}$ from the chain to 
the leads.
A particularly interesting property of the BALDA is its discontinuity at integer
values of the occupation number.\cite{LimaOliveiraCapelle:02} 
This discontinuity has a fundamental 
impact for time-dependent transport in the Coulomb blockade regime and may 
prevent a biased system from evolving towards a steady 
state.\cite{KurthStefanucciKhosraviVerdozziGross:10}

Finally, we note that for our Hubbard-like form of the interaction, where 
each electron interacts only with electrons of opposite spin on the same site, 
also the HF potential becomes a local potential depending only 
on the local density. In our numerical studies we will also present results 
obtained within the HF approximation.

\subsection{Adiabatic approximation: Steady-state condition for the density}
\label{steady-state}

Following the time evolution of the system as it is driven out of equilibrium 
by applying a bias in the leads, tells us if and how the system attains a 
steady state in the long-time limit. However, without doing the actual time 
propagation, we can also {\em assume} that the system approaches a steady 
state and study the consequences of this assumption. In the context of 
multistability in TDDFT, a particularly useful consequence is that, within 
the local and adiabatic approximation, it is possible to derive a 
self-consistency condition for the steady-state density.

Using non-equilibrium Green function techniques, the steady-state density  
$\tilde{n}_{j}:=\lim_{t \to \infty}n_{j}(t)$, can be obtained from the lesser 
Green function projected onto the central region, $\vG^<_{CC}$, as 
\be
\tilde{n}_{j} = 2\int \frac{{\rm d}\omega}{2\pi i}\; \left[ \vG^<_{CC}(\omega) 
\right]_{jj} \; .
\label{sceqn}
\ee
The lesser Green function can, in turn, be obtained from
\be
\vG^<_{CC}(\omega) = \vG^R_{CC}(\omega) \vSi^<_{\rm em}(\omega) 
\vG^A_{CC}(\omega) ,
\label{gcc<}
\ee
where $\vSi^<_{\rm em}(\omega)$ is the lesser embedding self-energy
 while $\vG^{R}_{CC}=[\vG^{A}_{CC}]^{\dag}$ is the retarded
KS Green function which, at the steady state, reads 
\be
\vG^{R}_{CC}(\omega) = \left( \omega  - 
\tilde{\vH}_{CC}^{KS}(\tilde{n}) - \vSi^R_{\rm em}(\omega) 
\right)^{-1}.
\label{ksgreen_ret}
\ee
In this formula 
\be
\tilde{\vH}_{CC}^{KS}(\tilde{n}):=\lim_{t \to \infty} 
\vH_{CC}^{KS}(t),
\ee
is the asymptotic value of the KS Hamiltonian in the central region.
Note that $\tilde{\vH}_{CC}^{KS}$ depends on $\tilde{n}$ only in the 
local and adiabatic approximation. 

For the 1D tight-binding leads the retarded embedding self-energy  
has the structure
\be
[\vSi_{\rm em}^R(\omega)]_{ij} = \Sigma_{{\rm em},L}^R(\omega) \delta_{i,1} 
\delta_{j,1} + \Sigma_{{\rm em},R}^R(\omega) \delta_{i,N_c} 
\delta_{j,N_c} , 
\label{emb_self} 
\ee
and therefore the self-consistency condition (\ref{sceqn}) becomes
\bea
\tilde{n}_{j} &=& \int \frac{{\rm d}\omega}{\pi} \left( f_L(\omega) 
\Gamma_L(\omega) |[\vG^{R}_{CC}(\omega)]_{1,j}|^2 \right. \nn\\ 
&& \left.+ f_R(\omega) 
\Gamma_R(\omega) |[\vG^{R}_{CC}(\omega)]_{N_c,j}|^2 \right),
\label{selfcons_dens_dft}
\eea
with the shifted Fermi function $f_{\a}(\w)=f(\w-\tilde{W}_{\a})$ 
and $\tilde{W}_{\a}:=\lim_{t \to \infty}W_{\a}(t)$. 
Since the KS Green function of Eq.~(\ref{ksgreen_ret}) depends on 
$\tilde{n}$ only, Eq.~(\ref{selfcons_dens_dft}) is a set of coupled nonlinear 
equations for the steady-state density $\tilde{n}_{j}$, 
$j=1,\ldots,N_C$, at the sites of the device region. Due to the nonlinearity 
of these equations, more than one fixed point solutions may exist. Therefore, 
one can use Eq.~(\ref{selfcons_dens_dft}) to scan the parameter space 
for possible multiple steady states. 
Once these values are identified we can investigate if and 
how the different steady states are attained by time propagation.

We close this Section by recalling a few basic properties of a fixed-point(FP) 
solution \cite{Kuznetsov:95} which we will use in the following to interpret 
and understand our numerical results. Consider a system of $N_{C}$ coupled 
equations of the following general form
\be
\vn=\bo{g}(\vn),~~\vn=(n_1,n_2,..., n_{N_C}) .
\label{fixedpoint}
\ee
Let the vector $\bo{\tilde{n}}$ be a fixed point solution of 
Eq.~(\ref{fixedpoint}) and $\bo{J}$ the Jacobian
matrix $\frac {d\bo{g}}{d\vn}|_{\bo{\tilde{n}}}$. 
A fixed point is stable if the modulus of all eigenvalues of $\bo{J}$ is 
smaller than unity, partially unstable if there exist at least one eigenvalue 
with modulus larger than unity (saddle point), and it is totally unstable if 
the modulus of all eigenvalues is larger than unity.

\section{Many-body technique: Kadanoff-Baym equations}
\label{mbpt}

An alternative approach to TDDFT for transport is the time-dependent MBPT 
formulation of transport  
\cite{MyohanenStanStefanucciLeeuwen:08,MyohanenStanStefanucciLeeuwen:09}
based on the time evolution of the Green function via the KB
equations.\cite{KadanoffBaym:62,dvl.2007,sdvl.2009,dahlen06procNGFT,dahlen06procKB} 
In the MBPT, the many-body self-energy $\Sigma_{\rm MB}$ can be approximated by a selection
of suitable Feynman diagrams, relevant for the description of the main scattering processes.
 In earlier work 
\cite {MBTDDFT:11}  we have found that the 2B approximation is in excellent agreement with accurate TD-DMRG results
in the regime of weak to intermediate interaction strength for the Anderson impurity model.\cite{HeidrichMeisnerFeiguinDagotto:09} 
Hence the 2B approximation is extremely useful for benchmarking other
approximations.
The  main quantity of MBPT is the Keldysh 
Green function, $\vG(z,z')$, where $z$ and $z'$ are time coordinates on the 
Keldysh contour ${\cal{C}}$.\cite{keldysh,danielewicz1,robert_06}
To describe the electron dynamics of the system, the Keldysh Green function 
is propagated in time according to the KB equations.
Since we are interested in the dynamical processes occurring in the central 
region, we can again use embedding techniques to derive the equation of motion 
for the Green function $\vG_{CC}$ projected on the central region. This 
equation reads 
\bea
\lefteqn{
\left[ i \partial_z - \vH_{CC}(z) \right] \vG_{CC}(z,z') = \delta(z,z')} \nn \\
&& + \int_{{\cal{C}}} {\rm d} \bar{z} \; \left[ \vSi_{\rm em}(z,\bar{z}) + 
\vSi_{\rm MB}(z,\bar{z}) \right] \vG_{CC}(\bar{z},z') \, ,
\label{eom_green_emb}
\eea
where, in addition to the embedding self-energy $\vSi_{\rm em}(z,\bar{z})$, we 
also have included a many-body self-energy $\vSi_{\rm MB}(z,\bar{z})[\vG]$ .
This later quantity is a functional of the Green function and 
fulfills all basic conservation laws.\cite{BaymKadanoff:61,Baym:62}  
We consider only the central region to be interacting and the leads are effectively noninteracting.
Therefore the many-body self-energy has nonvanishing elements only for the central region, because the diagrammatic
expansion starts and ends with an interaction line.
\cite{MyohanenStanStefanucciLeeuwen:08,MyohanenStanStefanucciLeeuwen:09}

\begin{figure}[t]
	\begin{center}
		\includegraphics[width=0.48\textwidth]{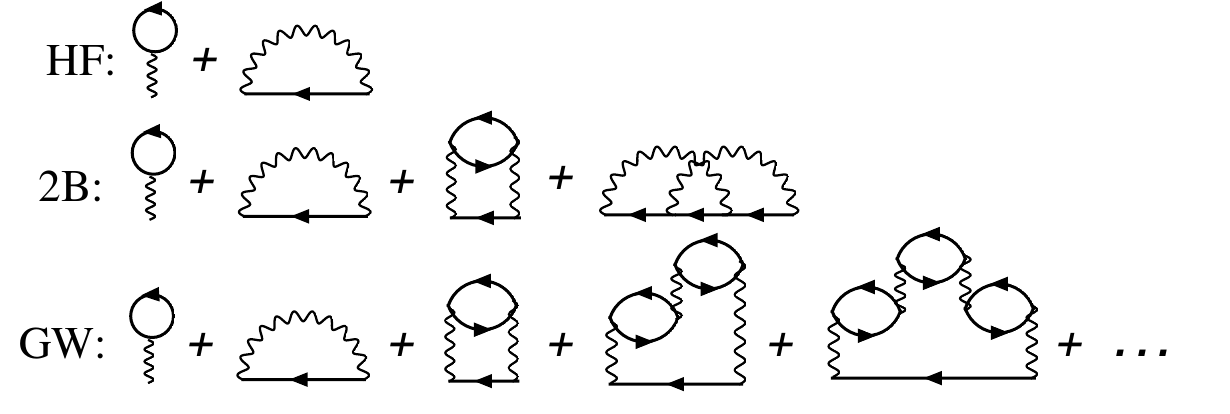}
		\caption{Conserving many-body approximations.}
		\label{diagrams}
	\end{center}
\end{figure}

Equation~(\ref{eom_green_emb}) is an exact equation for $\vG_{CC}$, provided an 
exact expression for $\vSi_{\rm MB}$ is inserted. Of course, for practical 
calculations the many-body self-energy must be approximated. In this 
paper we explicitly considered the following {\em conserving} 
approximations for the self-energy; the HF, 2B and $GW$ approximations. 
Their diagrammatic representations are illustrated in 
Fig.~\ref{diagrams}. 
The implementation of Eq.~(\ref{eom_green_emb}) requires a 
transformation into equations for real times, known as KB equations, which 
are then solved by time propagation
.\cite{dvl.2007,MyohanenStanStefanucciLeeuwen:08,MyohanenStanStefanucciLeeuwen:09,FriesenVerdozziAlmbladh:09}
From the knowledge of the Green function any one-particle property of the 
system can be extracted. In particular, the time-dependent density can be 
obtained from the lesser Green function as  
\begin{equation}
 n_j(t) =-2 i[\vG_{CC}^<(t,t^+)]_{jj} \; .
\label{tddensg<}
\end{equation}
In the correlated case there is no simplification such as Eq.~(\ref{dens_c}) since 
there are no more well defined single-particle states. Also
the current through lead $\alpha$ can be expressed in terms of 
the Keldysh Green  functions and reads\cite{MyohanenStanStefanucciLeeuwen:09,MyohanenStanStefanucciLeeuwen:08}
\begin{eqnarray}
\label{eq:current}
  \textrm{I}_{\alpha}(t)&=&2\te{Re}\bigg\lbrace \te{Tr}_{C}\Big[ \int_{t_0}^{t}d\bar{t}
 [\vG_{CC}^<(t,\bar{t})\vSi_{\rm em,\alpha}^A(\bar{t},t)\nonumber\\
&+&\int_{t_0}^{t}d\bar{t}\vG_{CC}^R(t,\bar{t})\vSi_{\rm em,\alpha}^<(\bar{t},t)]\nonumber\\
 &-i&\int_0^{\beta}d\bar{\tau}\vG_{CC}^{\rceil}(t,\bar{\tau})\vSi_{\rm em,\alpha}^{\lceil}(\bar{\tau},t)\Big]\bigg\rbrace.
\end{eqnarray}
Besides the superscripts $A$ (advanced), $R$ (retarded) and $<$ 
(lesser) we also introduced the superscripts $\lceil$ and $\rceil$ to 
denote the components with one time argument on the imaginary axis and 
the other on the  real axis and vice versa.\cite{StefanucciAlmbladh:04-2,robert_06} 
The trace is taken over one-electron indices in the central region. 
Eq.~(\ref{eq:current}) generalizes the Meir-Wingreen formula 
\cite{MeirWingreen:92} as it includes  initial  many-body and embedding 
effects through the integral along the imaginary track of the contour 
(last term).

For the interpretation of the numerical results we found it useful to 
look at the TD spectral function defined according to
\begin{equation}
\label{eq:spectral}
A(T,\omega)=- \textrm{Im}\, \textrm{Tr} \int\frac{\textrm{d}\tau }{\pi } 
e^{i\omega\tau}\big[ \vG^>-\vG^<\big](T+\frac{\tau}{2},T-\frac{\tau}{2}),
\end{equation}
where $\tau=t-t'$ is a relative time and $T=(t+t')/2$ is an average time 
coordinate. In equilibrium, this function is independent of $T$ and  
has peaks below the Fermi level at the electron removal energies 
and above the Fermi level  at the electron addition energies. If the 
time-dependent external field becomes constant after some switching time, 
then also the spectral function becomes independent of $T$ after some 
transient period and has peaks at the addition and removal energies of the {\em 
non-equilibrium} biased system.\\
 In the HF approximation the 
many-body self-energy is frequency independent and therefore the only 
broadening of the spectral peaks comes from the embedding part. 
This is also the case for the ABALDA.
When going beyond  the mean-field level, the self-energy becomes frequency 
dependent and as a consequence the peaks of the spectral function are typically broadened.

In the local and adiabatic approximation of TDDFT we have found a shortcut 
to determine if, for a given parameter set, the biased system can have 
multiple steady states (see Section \ref{steady-state}). It is important to note that a similar shortcut does 
not exist in MBPT. The reason is that in MBPT the self-consistency 
condition for the steady-state density requires the knowledge of 
$\vG_{CC}(\omega)$ (to compute the many-body self-energy 
$\vSi_{\rm MB}[\vG_{CC}]$) and thus the equations cannot be closed in 
terms of the density only. 
As a consequence, in MBPT one does not have an efficient method to 
scan the parameter space to look for bistability. 
Although this feature is somewhat inconvenient for our analysis, 
physically it is quite reasonable because it implies that possible 
dynamical XC effects (arising from the frequency dependence  
of the many-body self-energy) are included.   

\section{Results}
\label{result}
In this Section, we present the results of our numerical simulation 
for a certain set of parameters for which the self-consistency condition 
(\ref{selfcons_dens_dft}) admits multiple solutions.
We investigate  how one can switch
between different  stable solutions by applying a time-dependent gate voltage. 
We also demonstrate that for the same parameter sets the bistability 
is suppressed in the correlated many-body approximations, e.g., 2B and $GW$. 
The analysis will be carried out in two types of model devices, namely the one 
and two-site Hubbard models.

\subsection{Single site Hubbard model}

\begin{figure}[t]
	\centering
       \includegraphics[width=1.05\linewidth]{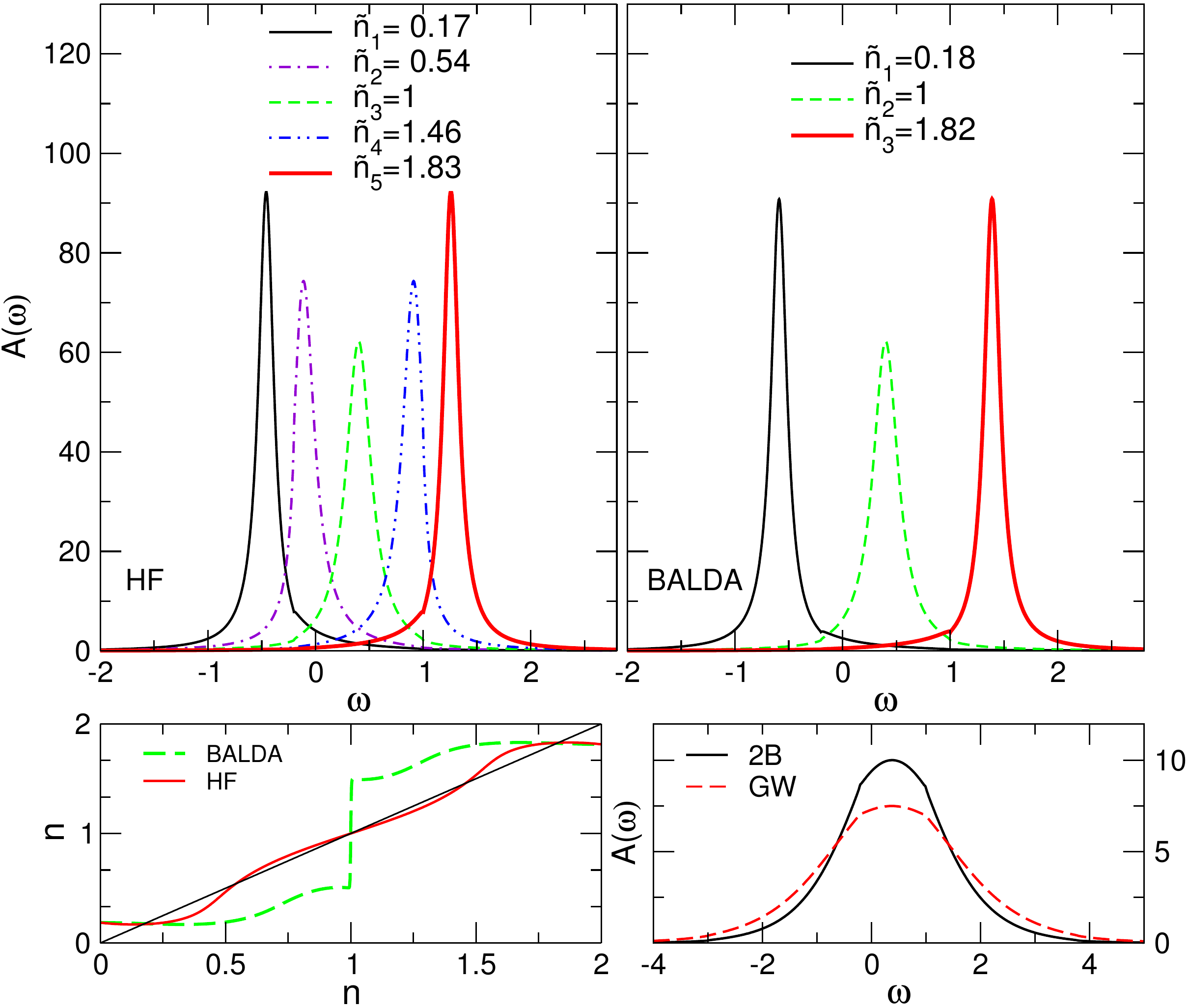}
       \caption[]{Spectral functions for the different steady-state 
       solution of the HF approximation (top-left panel) and BALDA (top-right 
       panel). The graphical solution of Eq.~(\ref{selfcons_dens_dft})
       is displayed in the bottom-left panel for the HF and BALDA. 
       For comparison we also report  the 2B and $GW$ steady-state spectral functions 
      in the bottom-right panel.}
      \label{fig:spectrum_five}
\end{figure}

 As a first example, we  study a single-site Hubbard model connected to semi-infinite
leads with the following parameters:
$V_{\rm link}$ = 0.3, $W_{L} = 1.8$, $W_{R} = -1.0$, $U$ = 2.0, $\varepsilon^{C}$ = -0.6, $\varepsilon_{\alpha} = 
\varepsilon_F=0$ (half-filled leads), and the inverse temperature 
$\beta$ = 90. All energies are measured in units of the lead-hopping 
parameter $V$. In the biased system the band-width of the leads extends from 
$\varepsilon_F+W_{\alpha}-2V$ to $\varepsilon_F+W_{\alpha}+2V$.
With these parameters the self-consistent  equation 
(\ref{selfcons_dens_dft}) admits 
five (three) solutions within the HF(BALDA) approximation. The fixed points are shown in the lower  
left corner of Fig.~\ref{fig:spectrum_five} where we 
display the left and right hand side 
of Eq.~(\ref{selfcons_dens_dft}). The corresponding densities 
for the HF are $\tilde{n}_1=0.17$, $\tilde{n}_2=0.54$, 
$\tilde{n}_3=1.0$, $\tilde{n}_4=1.46$ and $\tilde{n}_5=1.83$ while
for the BALDA the three fixed point densities are $\tilde{n}_1=0.18$, $\tilde{n}_2=1.00$, $\tilde{n}_3=1.82$. \\

For a single site the fixed point theorem tells us that  a solution 
 is stable if $|\frac{dg}{dn}| < 1 $,
with $g$ being the right-hand side of Eq.~(\ref{selfcons_dens_dft}). Hence, one can see
from Fig.~\ref{fig:spectrum_five} that the fixed points 
$\tilde{n}_1$, $\tilde{n}_3$ and $\tilde{n}_5$ are stable in the case 
of the HF, while  
in case of the BALDA the stable solutions are $\tilde{n}_1$ and 
$\tilde{n}_3$. Although the solution with density of unity   
exists for both approximations, it is
stable in the HF approximation and unstable in the BALDA.
In the upper panels of Fig.~\ref{fig:spectrum_five} we plot the steady-state spectral 
functions corresponding to the fixed points of the HF and BALDA.
The HF peak for density $\tilde{n}_1 = 0.17 $  (and the BALDA peak 
for density $\tilde{n}_1 = 0.18$) is located
within the right lead energy continuum, while the HF peak for density
$\tilde{n}_5 = 1.83$ (and the BALDA peak for density $\tilde{n}_3 = 1.82$) is 
located within the left lead energy continuum.
By contrast, the HF spectral function of the unstable fixed points,  
$\tilde{n}_2$ and $\tilde{n}_4$, are peaked at the edges of left and right lead 
band respectively.
The HF and  BALDA spectral functions of the
fixed point with density of unity are identical (the XC potential is 
zero in this case) and the peak is 
located  exactly in the middle of the overlapping region between the 
left and right bands.
In spite of this the stability condition of this fixed point is 
completely different in the HF and BALDA case.  
Since the multistability  can be most easily observed if 
the spectral peaks of the stable solutions are well separated, we 
conclude that this phenomenon is favoured when
the  energy bands have a small overlap and the system is in the   
negative differential resistance (NDR) regime.
\begin{figure}[t]
  \centering
  \includegraphics[width=1\linewidth]{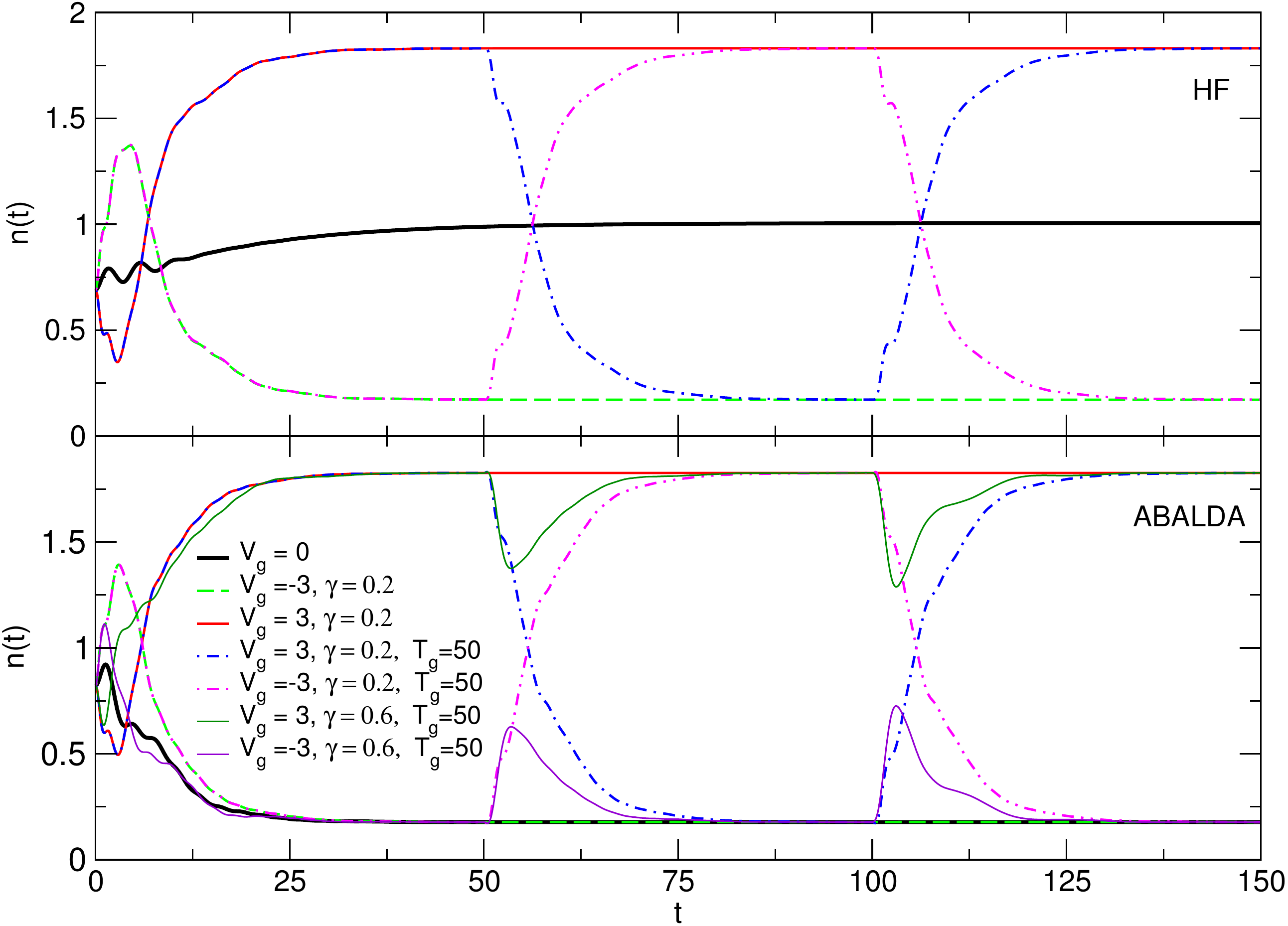}
  \center{\includegraphics[width=1\linewidth]{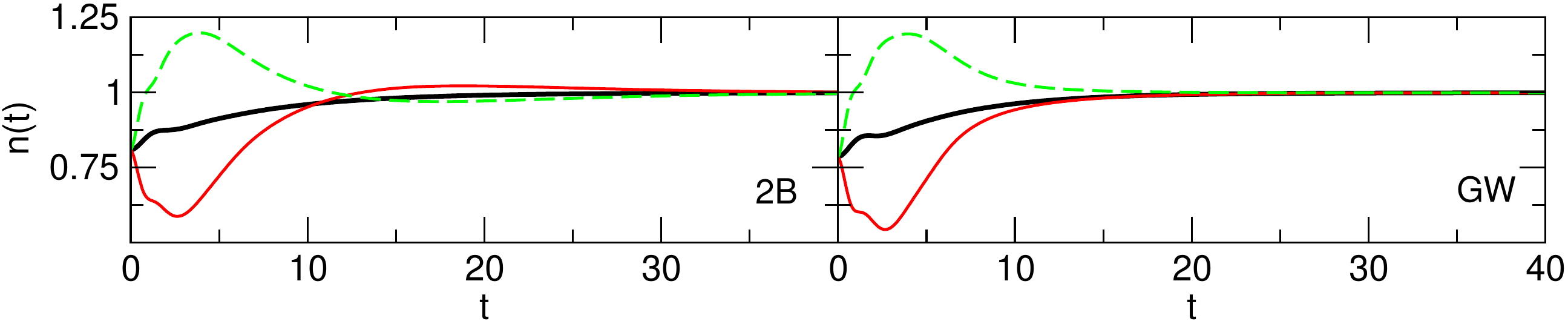}}  
  \caption[]{Top panel: Time-dependent density in the HF approximation (top) and ABALDA 
  (bottom) after the sudden switch on of the bias voltage and 
  a series of gate pulses as in Eq.~(\ref{eq:gate}). Bottom panel: 
  Time-dependent density within 2B (left) and $GW$ approximations (right) after the 
  sudden switch on of the bias voltage and a gate pulse as in Eq.~(\ref{eq:gate}) 
  with $V_{g}=-3,0, 3$.}
  \label{fig:density_five_all}
\end{figure}
As we shall see below, for the correlated MBPT approximations the situation is very different.
In the lower left panel of Fig.~\ref{fig:spectrum_five} we show the 
2B and $GW$ steady-state spectral functions,
as obtained from the propagation of the KB equations. 
The spectral weight
is spread over the whole lead energy range and beyond. 
Consequently, the height of the spectral
function is also much smaller. The considerable broadening is due to 
an increased quasi-particle scattering in the out of equilibrium 
system as already observed in Ref. \onlinecite{MyohanenStanStefanucciLeeuwen:09}. 

Let us now study  how to switch between different stable steady-state 
densities using ultrafast time-dependent driving fields. We start 
from the initially unbiased equilibrium system with ground-state density  
$\tilde{n}_0=0.69 $ ($\tilde{n}_0=0.82$)  for HF (BALDA).
 In Fig.~\ref{fig:density_five_all} we show the time evolution of the density at 
the interacting site for different approximations after the sudden 
switch on of the bias voltage $W_L=1.8$ and  $W_R=-1.0$.
In the HF approximation we observe that  after some transient  time the
density approaches the value 1. The behavior of the ABALDA density is 
very different, in agreement with the fact that the solution with 
density of unity is  unstable and hence cannot be reached 
by  time evolution. At the steady state the ABALDA density equals 
the lowest value  $\tilde{n}_1$. 

\begin{figure}[tpb]
	\centering
       \includegraphics[width=\linewidth]{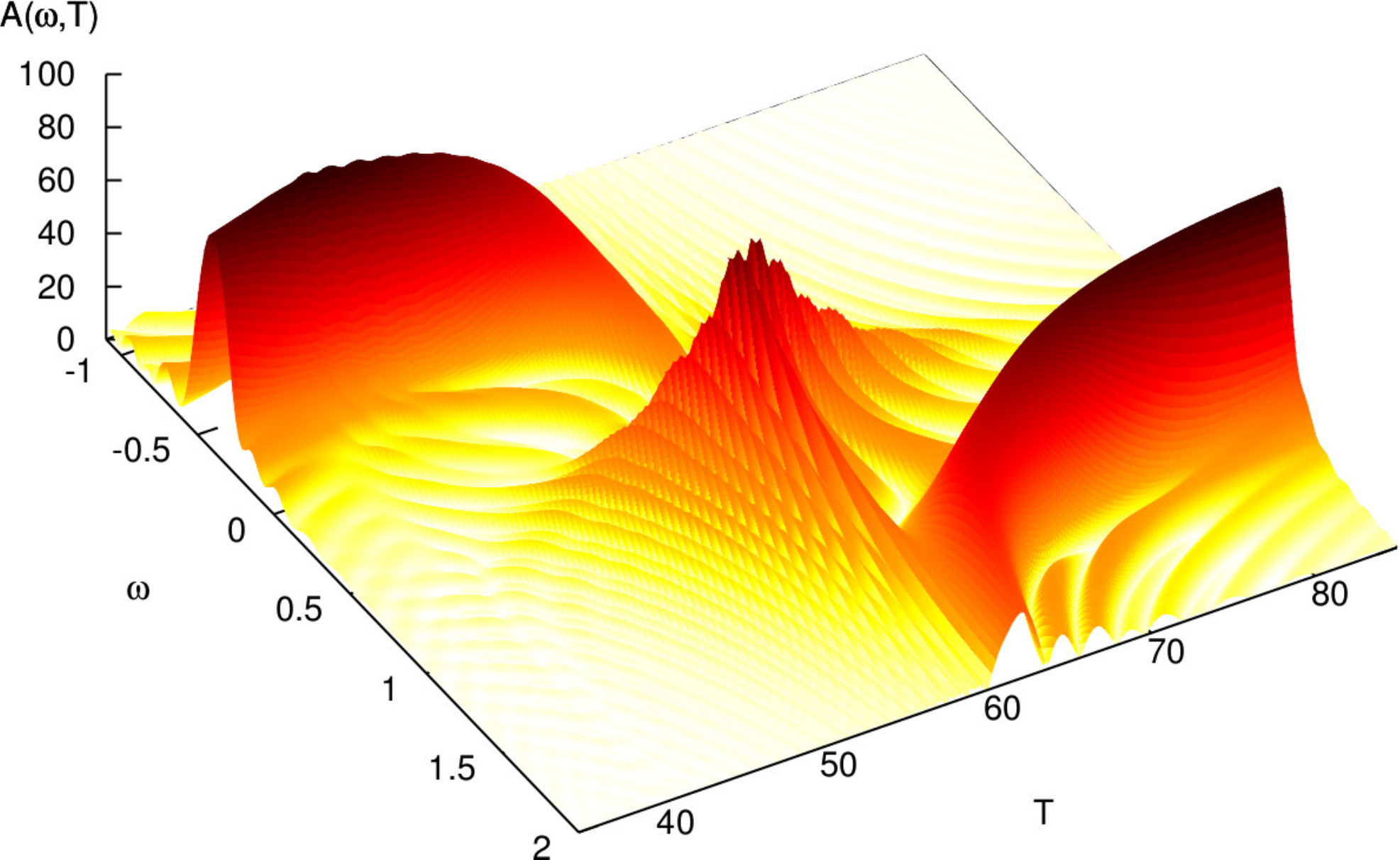}
       \caption[]{Non-equilibrium spectral function for a gate pulse
$V_g=-3$, $\gamma=0.2$ which brings the density to 
$\tilde{n}_1$ first, followed by a second identical gate pulse gate 
but with opposite amplitude which brings the density to $\tilde{n}_5$.
The intermediate transition to the $\tilde{n}_3$  stable solution is 
clearly visible.}
\label{fig:noneq_spec}
\end{figure}

To switch to the other stable solutions in real time we applied a 
time-dependent gate pulse on the Hubbard site. 
In this work we have used an exponentially decaying gate voltage of the form
\begin{equation}
\label{eq:gate}
 V_g(t)=\begin{cases}
        V_ge^{-\gamma t}         & ,\;\textrm{if}\;\;0<t<T_{g}     \\
        -V_ge^{-\gamma(t-T_{g})} & ,\;\textrm{if}\;\;T_{g}<t<2T_{g}\\
        V_ge^{-\gamma(t-2T_{g}) }&,\;\textrm{if}\;\;t>2T_{g}      \\
        \end{cases}.
\end{equation}  
In Fig.~\ref{fig:density_five_all} we show that 
in the HF case, the state with the lowest density $\tilde{n}_{1}$ can be obtained
(in addition to applying a sudden bias in the leads) by
switching on a pulse with amplitude $V_g =-3.0$, decay rate $\g =
0.2$ and $T_g=\infty$. The state with highest density $\tilde{n}_5 =
1.83$ can be obtained in a similar fashion but now applying a
gate with positive amplitude $V_g = 3.0$.
Thus, by changing the amplitude of the gate voltage we can switch between stable steady-state
solutions. For instance, with
a first pulse of positive amplitude and $T_{g}=50\gg 1/\gamma$
the system reaches the state with $\tilde{n}_{5}$.
At the time $T_g$ we apply a second pulse
but with negative amplitude. The density shows a transient
behavior after which it approaches the value 
$\tilde{n}_1$. If we now apply a 
third pulse of positive amplitude at time $2 T_g$  the density goes back  
to the initial value $\tilde{n}_{5}$. This is nicely illustrated in 
Fig.~\ref{fig:density_five_all}.
In Fig.~\ref{fig:noneq_spec} we show
the non-equilibrium HF spectral function $A(T,\omega)$ of Eq.~(\ref{eq:spectral}) for a
double switch with $V_g=-3.0$, $\gamma=0.2$ and $T_{g}=50$. 
The figure enlightens an interesting aspect regarding the transition 
from one steady-state to another.
The density rises from the lowest $\tilde{n}_{1}$ 
to the highest $\tilde{n}_{5}$ lingering for a while in the middle stable 
solution $\tilde{n}_{3}$.

Going back to Fig.~\ref{fig:density_five_all} we see that also in the 
ABALDA the state with highest density $\tilde{n}_3 = 
1.83$ is reached by applying a
gate pulse with $V_g = 3.0$ and $\g=0.2$ (in addition to  a sudden bias in the leads).
If the amplitude is negative instead ($V_g = -3.0$) the 
density increases first but eventually 
drops down and goes back to its initial value $\tilde{n}_1$.  
Like in the HF case we
can switch back and forth between stable solutions by changing the 
sign of $V_{g}$. 
Not unexpectedly, however,
the decay time $\tau_{\gamma}\sim 1/\gamma$ cannot be arbitrarily short. 
If $\tau_{\gamma}$ is too short 
the system does not have time to accumulate or lose enough density to 
change the self-consistent potential
and after some transient it falls back to the previous steady-state 
value. This is clearly shown in
Fig.~\ref{fig:density_five_all} for the amplitude $V_g = \pm 3.0$, 
$T_g=50$ and a faster decay rate $\g =0.6$.

Intuitively one would expect that by increasing (decreasing) the on-site energy 
of the Hubbard site the  density decreases (increases). However, the highest (lowest) 
stable steady-state density  is obtained with a positive (negative) gate. This is due to 
the  fact that in our case the on-site energy of the Hubbard site 
lies below the energy  band of the left lead. By 
applying  a positive gate a finite hybridization occurs, leading to the migration of 
extra  charge from the left lead to the Hubbard site. A similar 
argument  explains the  reduction  of the density  on the impurity 
site  when a negative gate is turned on.
 
In the lower panels of Fig.~\ref{fig:density_five_all}  
we plot the densities obtained within the 2B and
$GW$ self-energy approximations. We applied the bias voltage and a 
gate pulse of the 
form $V_g(t)=V_ge^{-\gamma t}$ for $t >0$ with $V_{g}=0,\;\pm 3$. In 
all cases  only one steady state emerges at the end of the propagation
with a density of about $1.0$. 
It is worth observing that the 2B and $GW$ steady-state values of the densities are 
close to each other, indicating  that 
the single-bubble diagram, 
common to both approximations, is the dominant term of the 
perturbative expansion in this case.\cite{MyohanenStanStefanucciLeeuwen:09}

By time propagation we have shown that the three 
HF densities, $\tilde{n}_1$, $\tilde{n}_3$ and $\tilde{n}_5$, 
and the two  ABALDA densities, $\tilde{n}_1$ and $\tilde{n}_3$, are stable
in a slightly different sense than that of the fixed 
point theorem. The fixed point theorem does not  contain any information on the actual dynamics.
Similarly, the HF solutions $\tilde{n}_2$ and $\tilde{n}_4$ as well 
as the ABALDA solution $\tilde{n}_2$ are unstable in the sense that 
there exist no external perturbation to drive the system toward them.
Thus, the fixed point theorem provides us with a good 
criterion to establish whether a given steady-state can be reached or not.
This criterion is certainly rigorous in the limit of adiabatic 
switchings but, as we just found, its validity extends well beyond 
the adiabatic regime.
\begin{figure}[t]
	\centering
        \includegraphics[width=\linewidth]{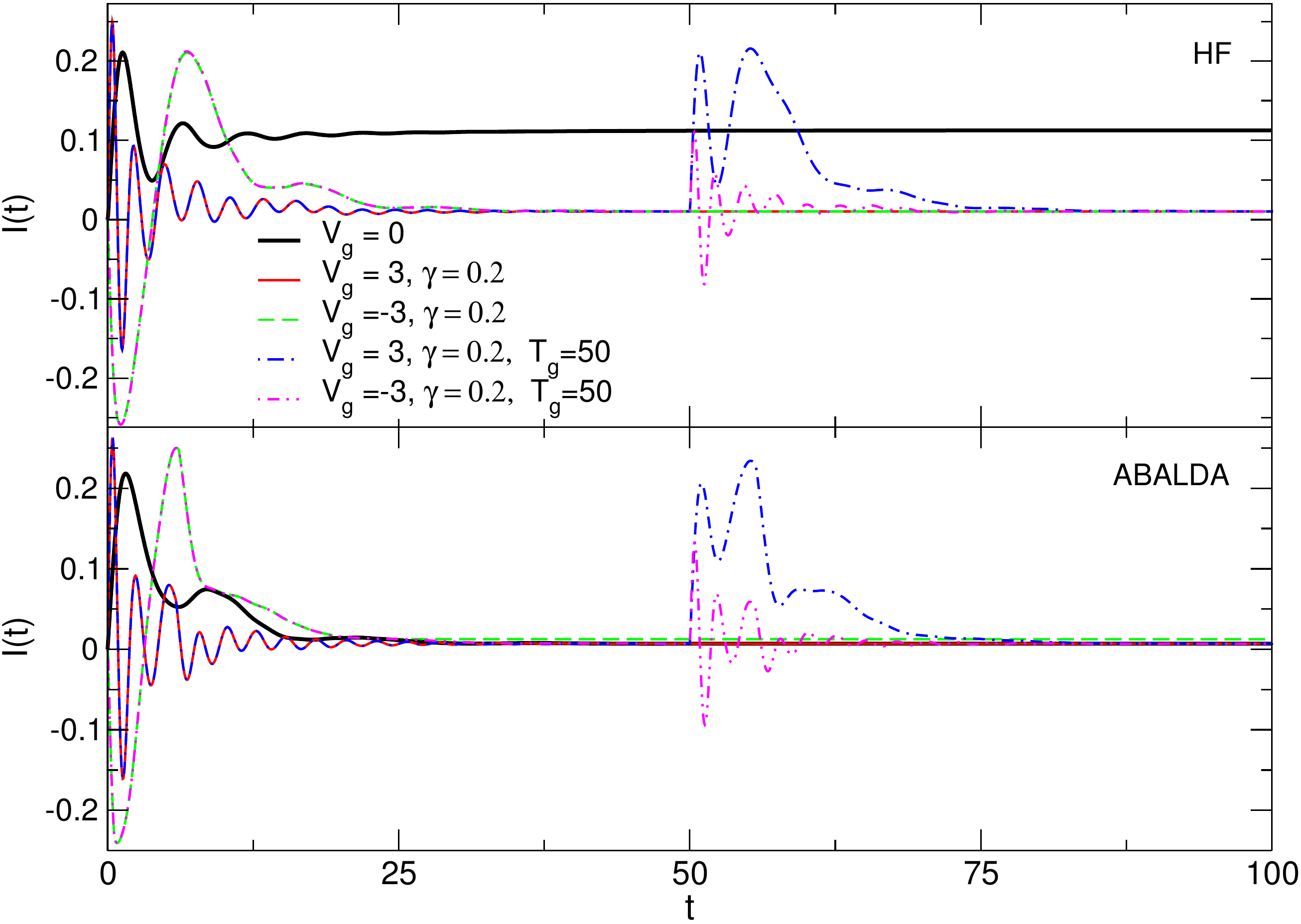}
        \includegraphics[width=0.99\linewidth]{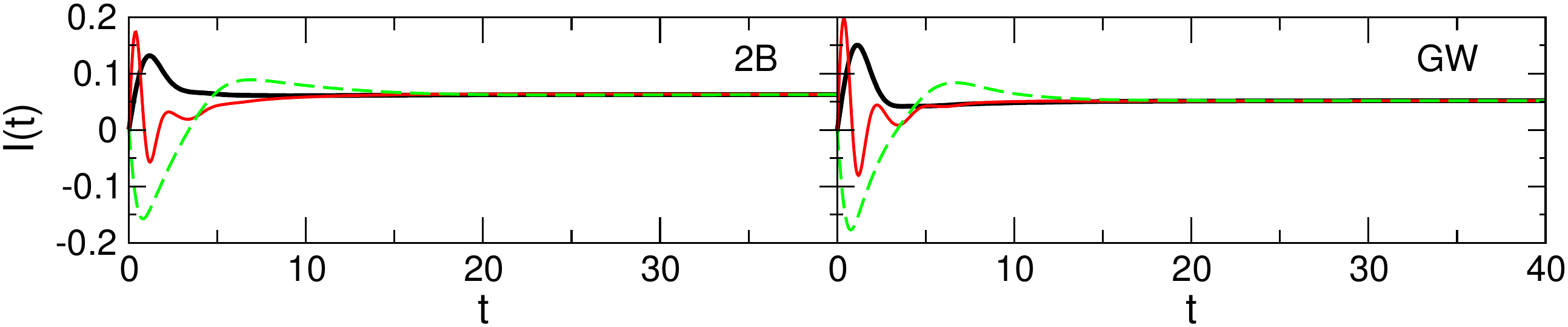}
        \caption[]{Time-dependent current after the sudden switch on of 
	the bias voltage and of a gate pulse as in 
        Eq.~(\ref{eq:gate}) in the HF (top-upper panel), ABALDA 
	(top-lower panel), 2B (bottom-left panel) and $GW$ 
	(bottom-right panel) approximations. In the HF and ABALDA case a time-dependent 
	switch between two different steady-states is shown.}
        \label{fig:current_five_all}
\end{figure}

The time-dependent currents for the various approximations 
are shown in Fig.~\ref{fig:current_five_all}. 
Corresponding to the three stable HF steady-state
densities there exist only two distinguishable values
for the current $I_R(t)$ at the interface between 
the Hubbard site and the right lead. The
lower value corresponds to the solutions $\tilde{n}_1$ and $\tilde{n}_5$, while the higher value 
corresponds to the solution $\tilde{n}_3$. 
The existence of only two solutions
for the current is the consequence of an
approximate particle-hole symmetry of the self-consistent equation 
(\ref{selfcons_dens_dft}), i.e., $\tilde{n}_{5} \sim 1-\tilde{n}_{1}$. 
One particular appealing feature of the HF currents is the large 
difference  between the two steady-state values, a  highly  
desirable property for designing nanoscale diodes.    

The particle-hole symmetry
holds also for the ABALDA and therefore the steady-state currents corresponding to the 
two stable solutions are almost indistinguishable. 
Finally, the 2B and $GW$ steady-state values of the currents, approach the same value 
independent of the gate voltage, in agreement with the existence of a 
unique steady state. 

Increasing the bias the number of HF fixed point solutions reduces to three of 
which only two are stable. 
Also by increasing
the interaction strength $U$ the number of stable solutions 
reduces to two because a small amount of density 
causes a considerable change in the effective potential.
Consequently, the middle solution becomes unstable. 

\subsection{Two site Hubbard model}
In this Section we consider the case of two interacting 
sites ($N_C=2$) connected to two semi-infinite, non-interacting tight-binding 
leads. We choose the following parameters:
$V_{\rm link}$ = 0.4, $W_{\textsc{L}}$ = 2.2,  $W_{\textsc{R}}$ = -1.2, $U$ = 2.0, 
$V_C = V_{1,2}$ = 0.4, $\varepsilon_{\alpha} = \varepsilon_F=0$, 
$\varepsilon^{C}_{1} = \varepsilon^{C}_{2}$= -0.6 and $\beta$ = 90.
The leads are half-filled and the lead bands have an energy range between 
$\varepsilon_{\alpha}+W_{\alpha}-2V$ and $\varepsilon_{\alpha}+W_{\alpha}+2V$. 

\begin{figure}[t]
 \begin{center}
   \center{\includegraphics[angle=0, width=1.05\linewidth]{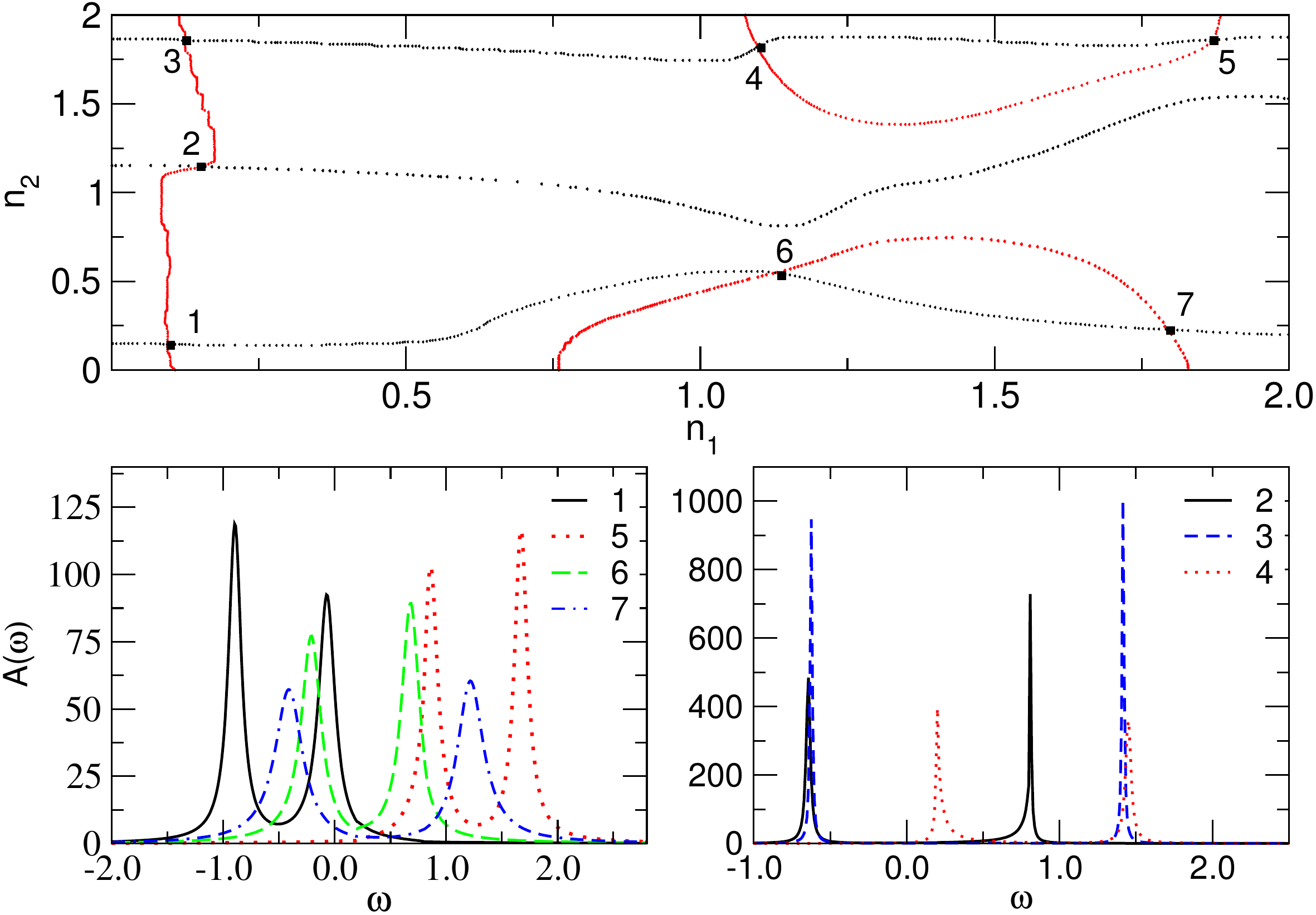}}
    \caption[]{ Upper panel: Graphical solution of the integral in Eq.~(\ref{selfcons_dens_dft}).
     Lower panel: Spectral functions for the HF approximation with Hubbard interactions
     corresponding to the seven different steady-state solutions for 
     the density.}
   \label{fig:steady_HF_dots}
 \end{center}
\end{figure}

Within the HF approximation, the steady-state condition (\ref{selfcons_dens_dft}) then 
has seven solutions which are shown in the upper panel of 
Fig.~\ref{fig:steady_HF_dots}. The black curve is obtained by finding the
root of the equation $n_2-g_2(n_1,n_2)=0$ at fixed $n_1$ where 
$g_2(n_{1},n_{2})$ is the right hand side of Eq.~(\ref{selfcons_dens_dft}) with $j=2$. 
The red curve is obtained in an analogous way by exchanging $1\leftrightarrow 2$. 
Hence the intersections of the curves are the fixed points. 
The numerical values for the steady-state densities at the two Hubbard sites for the 
seven fixed points are given in Table \ref{table_HF_fix}. 

\begin{table}
\begin{tabular}{|c|c|c|c|c|c|}
\hline
FP & $n_1$ & $n_2$ & FP & $n_1$ & $n_2$ \\ \hline
1 & 0.094 & 0.144 & 5 & 1.867 & 1.862 \\
2 & 0.150 & 1.146 & 6 & 1.129 & 0.546 \\
3 & 0.124 & 1.860 & 7 & 1.794 & 0.226 \\
4 & 1.098 & 1.821 &  & & \\ \hline
\end{tabular}
\caption{Fixed point (FP) solutions of Eq.~(\ref{selfcons_dens_dft}) for the 
steady-state densities of two interacting Hubbard sites connected to two biased, 
non-interacting leads in the HF approximation (see upper panel of 
Fig.~\ref{fig:steady_HF_dots}). 
The parameters are: $V_{\rm link}$ = 0.4, $W_{\textsc{L}}$ = 2.2,  
$W_{\textsc{R}}$ = -1.2, $U$ = 2.0, $V_{1,2}$ = 0.4, 
$\varepsilon_{\alpha} = \varepsilon_F = 0$, and  
$\varepsilon^{C}_{1} = \varepsilon^{C}_{2}= -0.6$.  }
\label{table_HF_fix}
\end{table}

In the lower panel of Fig.~\ref{fig:steady_HF_dots} we show the spectral functions
corresponding to the seven different fixed points (FP's). The spectral function for 
FP 1 is located mostly in an energy range within the energy band of the right lead, 
while the one for FP 5 has most of its weight in the energy range of the left lead. 
In contrast, FP's 6 and 7 have considerable weight in the energy bands of both leads.
The spectral functions corresponding to FP's 2, 3, and 4 have much narrower peaks 
than the spectral functions for the other fixed points. 

\begin{figure}[t]
	\centering
 
        \includegraphics[angle=0,height=0.31\textheight,width=1.0\linewidth,clip]{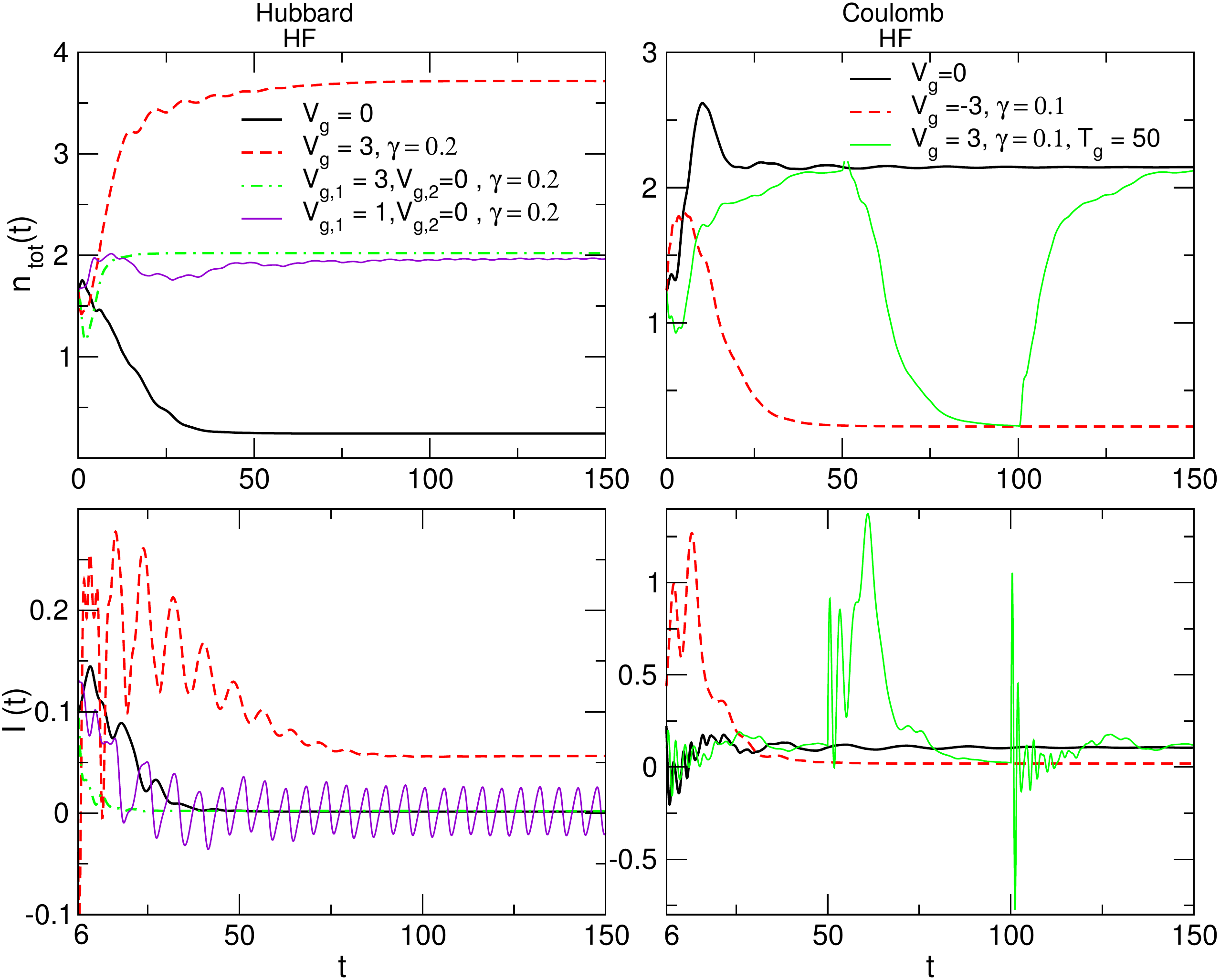}      
  \caption[]{Densities and currents for the HF approximation  in case of  short
       range (Hubbard) and long range (Coulomb) interactions.  A switch
       between different steady-states by applying exponentially 
       decaying gate of Eq.~(\eqref{eq:gate}) 
       are shown.
       }
      \label{fig:td_hf_dot}
\end{figure}

According to the fixed point theorem only the FP's 1, 3, 5, and 7 are stable and 
we expect them to be accessible by time propagation. In the upper left panel of 
Fig.~\ref{fig:td_hf_dot} we show the time evolution of the total density on the two 
dots, $n_{\rm tot}(t)=n_1(t)+n_2(t)$, in the HF approximation 
for a sudden switch-on of the bias and several gate voltages starting from the
equilibrium state with ground state density $n_1=n_2=0.83$. 
The steady state corresponding to FP 1 is obtained by applying only the bias (no gate).
In the case where we apply, in addition to the bias, a decaying gate voltage
of the form~\eqref{eq:gate} to both sites with $V_{g,1}=V_{g,2}=V_g=3.0 $ and 
decay rate $\g=0.2$, the total density increases and after 
some transient evolves towards the steady state corresponding to FP 5.
In this case, lifting the on-site energy due to the gate voltage allows for extra 
charge to accumulate at the interacting sites such that the high-density steady-state 
solution can be achieved. 
In contrast, the solution of FP 7 can be obtained by applying the decaying gate 
voltage to the first (left) site 
only with amplitude $V_{g,1}=3.0$, $V_{g,2}=0 $ and $\g=0.2$.

Surprisingly, applying a similar asymmetric gate voltage but with a smaller 
amplitude ($V_{g,1}=1.0, V_{g,2}=0$), leads to a very different long time 
behavior. In this case the system does not evolve towards a steady state after 
the transients, instead we observe an oscillatory time-dependent density.   
In the long-time limit, the time-dependent total density (purple curve in the 
upper left panel of Fig.~\ref{fig:td_hf_dot}) oscillates with an amplitude of 
the order of $10^{-3}$, around  $1.96$. This value corresponds approximately 
to the total steady-state density of FP 3 of Fig.~\ref{fig:steady_HF_dots}. 

\begin{figure}[t]
	\centering
          \includegraphics[angle=0,height=0.26\textheight,width=1.0\linewidth,clip]{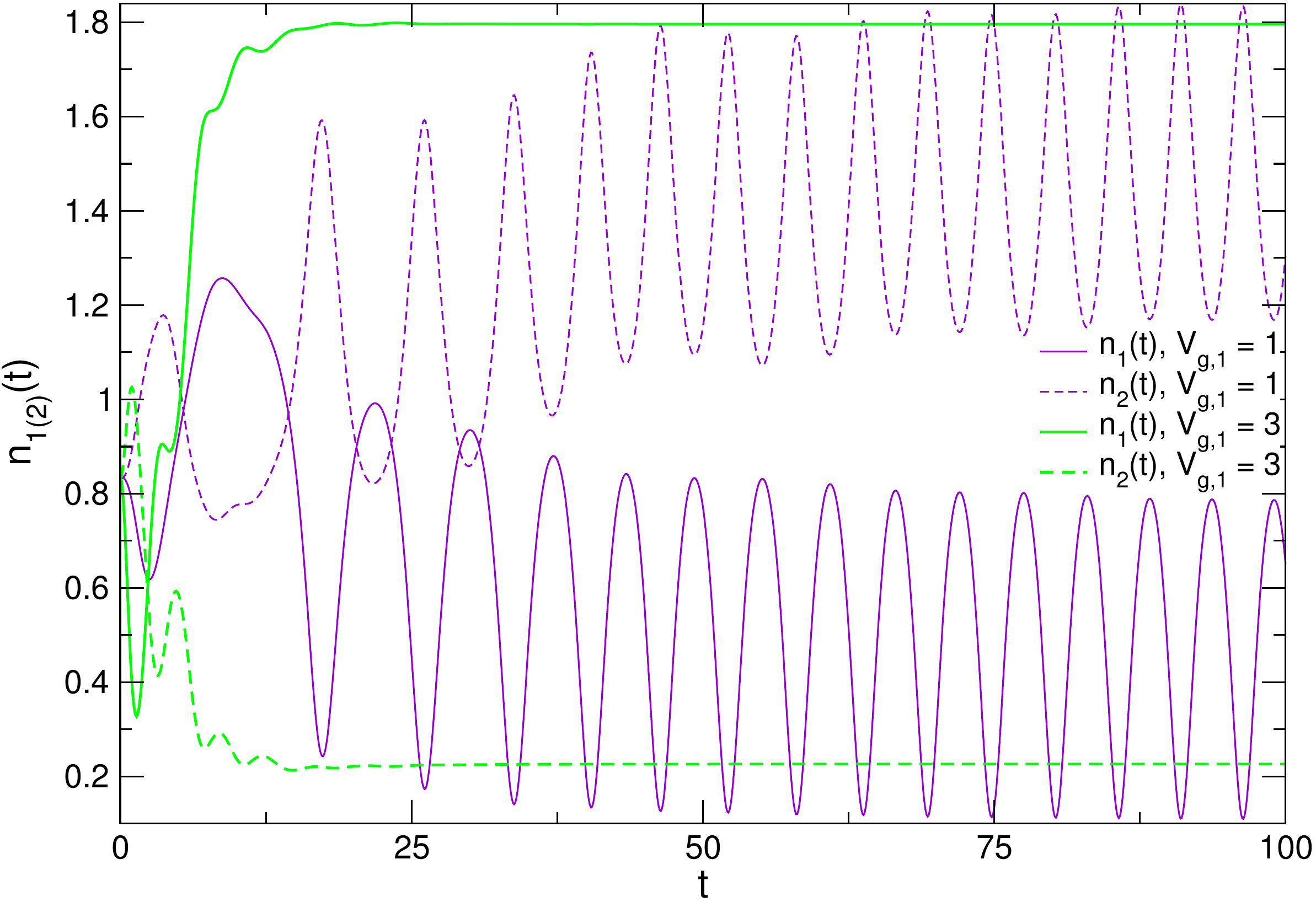}      
  \caption[]{Densities of the first and second site for the HF approximation  in case of  
       Hubbard interactions corresponding to the  middle curves in the upper 
       left panel of Fig.~\ref{fig:td_hf_dot}
       }
      \label{fig:td_dens12}
\end{figure}

Despite this apparent similarity, the nature of 
these solutions is very different. While for the steady state of FP 7 the 
charge is mostly located at the first site, in the case of FP 3 the density on 
the first site is smaller than on the second one (see 
Table~\ref{table_HF_fix}). The different nature of these two cases becomes even 
more obvious when looking at the time evolution of the density at the two 
interacting sites separately (see Fig.~\ref{fig:td_dens12}). While in the 
first case ($V_{g,1}=3.0$, $V_{g,2}=0 $) the steady state is attained quite 
rapidly, in the second case ($V_{g,1}=1.0, V_{g,2}=0$) we see non-decaying 
density oscillations at the individual sites with rather large amplitudes. 
In the long-time limit the density oscillates thus inducing an oscillating 
KS potential. The persistence of these oscillations means that the density 
solves the Floquet system of equations in which the harmonics of the 
potential depend on the density itself. 
At first sight one might be reminded of non-decaying density and current 
oscillations which can appear for non-interacting systems when the biased 
system possesses two or more bound states. 
However, here we work in the HF approximation and therefore 
the analysis of Refs. \onlinecite{Stefanucci:07,KhosraviStefanuccikurthGross:09,KhosraviKurthStefanucciGross:08}
needs to be modified, see below. 

\begin{figure}[t]
	\centering
          \includegraphics[angle=0,width=1.0\linewidth,clip]{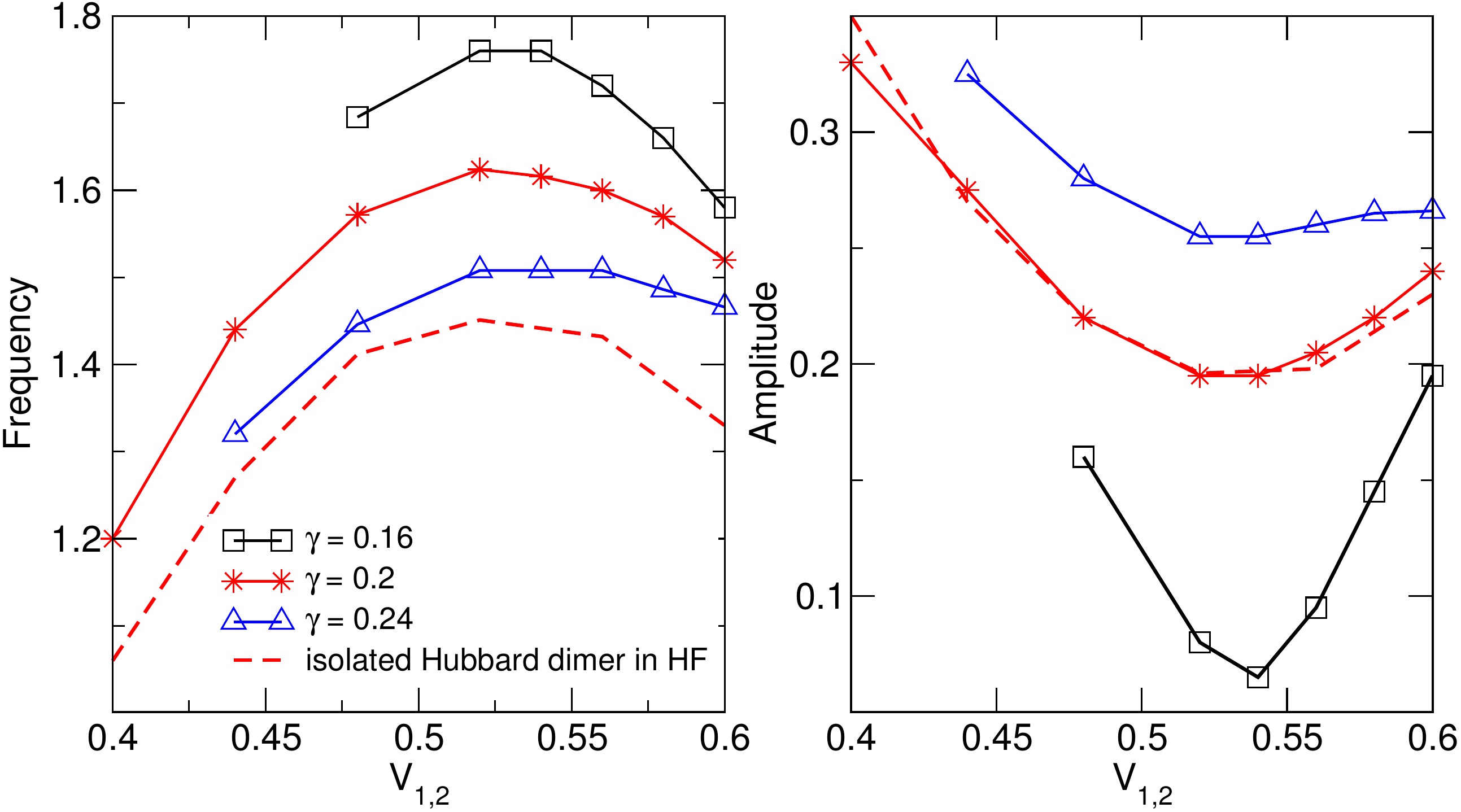}
  \caption[]{Oscillation frequency and amplitude of the density oscillations 
found in HF for certain gate switchings of the Hubbard dimer 
connected to biased leads as function of the hopping between the Hubbard 
sites. For comparison, oscillation frequencies and amplitudes are given for 
the isolated Hubbard dimer in HF approximation described by Eq.~(\ref{hf_dimer_dens}).}
      \label{fig:comp_osc_dimer}
\end{figure}

Some insight into the nature of these oscillations can be gained from the 
simple model of an {\em isolated} Hubbard dimer. In the HF 
approximation, the equation of motion for the electronic density matrix 
$\rho$ of the isolated dimer reads
\be
i \partial_t \rho(t) = [ H_{HF}(t),\rho(t) ] ,
\ee
where the HF Hamiltonian is given by 
\be
H_{HF}(t) = 
\left( \begin{array}{cc} \varepsilon_1 + U n_1/2  & V_{1,2} \\ 
V_{1,2} & \varepsilon_2 + U n_2/2  \\ 
\end{array} \right) \;.
\label{hf_dimer}
\ee
Under the simplifying assumption $\varepsilon_1=\varepsilon_2$ one can then 
derive the equation of motion for the quantity 
$\delta n(t)= n_1(t)-n_2(t)=\rho_{11}(t)-\rho_{22}(t)$ which reads
\be
\delta \ddot{n} + (4 V_{1,2}^2 -  U D) \delta n + \frac{U^2}{8} (\delta n)^3 
= 0 ,
\label{hf_dimer_dens}
\ee
where the constant $D$ is related to the initial condition of 
Eq.~(\ref{hf_dimer}) and can be defined through the off-diagonal matrix 
elements of the density matrix as 
\be
D = V_{1,2} (\rho_{1,2}(0) + \rho_{2,1}(0)) + \frac{U}{8} (\delta n(0))^2 \; .
\ee
We note that Eq.~(\ref{hf_dimer_dens}) is the equation of motion of a 
classical, anharmonic oscillator and therefore supports oscillating solutions. 
We now check if the model of the isolated Hubbard dimer has anything to do 
with the oscillations seen in our transport setup. To this end, we calculate
$D$ from Eq.~(\ref{hf_dimer_dens}),
i.e., $D=\frac{\delta \ddot{n}} {\delta n U}+ \frac {4 V_{1,2}^2}{U} +\frac{U}{8} (\delta n)^2$,
where the densities and their time-derivatives are taken from the transport 
calculation after the transients have died out.
As the Eq.~(\ref{hf_dimer_dens}) is an approximation for the connected Hubbard dimer, $D$ is not constant in time. 
Hence in order to compare the oscillation amplitudes and frequencies from the transport simulations
 with those resulting from Eq.~(\ref{hf_dimer_dens}) we averaged $D$ over an oscillation period. 
In Fig.~\ref{fig:comp_osc_dimer} we show the dependence of oscillation 
frequency and amplitude for different switchings of the gate ($V_{g,1}(t)=
V_0 \exp(-\gamma t)$, $V_{g,2}(t)=0$, $V_0=1$)
as function of the hopping $V_{1,2}$ between the two sites of the 
Hubbard dimer connected to biased leads and compare to the corresponding 
solutions of Eq.~(\ref{hf_dimer_dens}). We see that both frequency and 
amplitude of the isolated and connected dimer behave qualitatively quite 
similar as function of the intersite hopping and we conclude that the 
model of the isolated dimer certainly captures the physics behind 
these oscillations. 
We also would like to point out that the regions of parameter space where the 
oscillations are found appears to be quite small. For most parameters the 
system actually does evolve towards one of the steady states of 
Table \ref{table_HF_fix}. 

\begin{figure}[t]
	\centering
       \includegraphics[angle=0,height=0.3\textheight,clip]{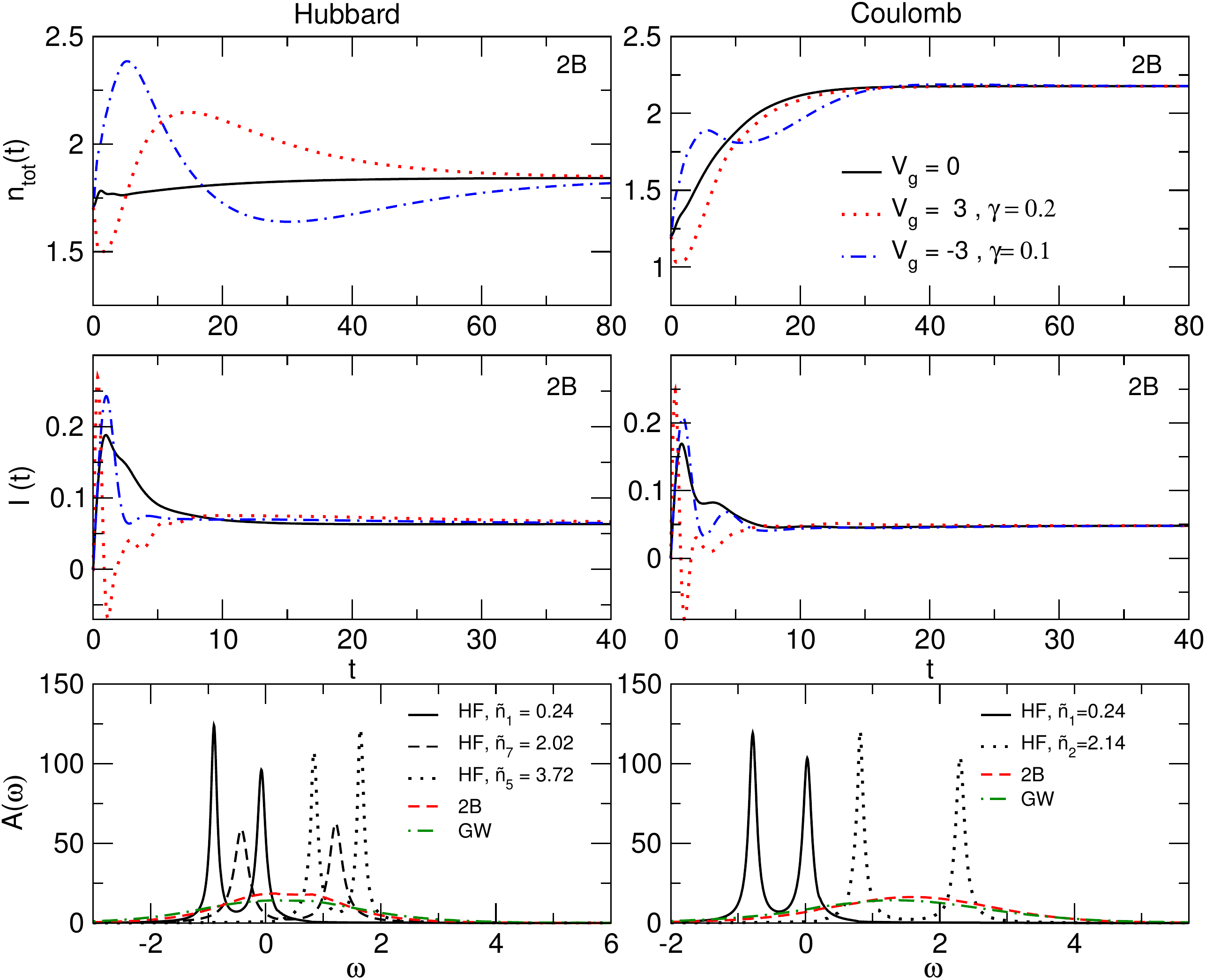}
       \caption[]{Densities and currents for the 2B approximation short
       range (Hubbard) and long range (Coulomb) interactions are used. {\it Lower panel}:
       Spectral functions for the different approximations at the end of the time propagation.}
      \label{fig:td_2b_dot}
\end{figure}

The occurrence of self-induced persistent oscillations in the HF mean 
field theory is favoured by the short-range nature of the 
 Hubbard interaction. In fact, we also have studied a modified version of our model where we 
replaced the last term of Eq.~(\ref{hamilc}) by a more long-range, 
Coulomb-like interaction $\frac{1}{2} \sum_{\substack{ i=1 \\\s\s'}}^{2} 
U_{i,j}\hat{d}_{i\sigma}^{\dagger} \hat{d}_{j\sigma'}^{\dagger} 
\hat{d}_{j\sigma'} \hat{d}_{i\sigma} $ with
\begin{equation}
\label{eq:coul-int}
U_{i,j}=\begin{cases}
U  & i=j     \\
\frac{U}{2|i-j|} & i\neq j \\
\end{cases}.
\end{equation} 
In this case we have not found any oscillating solutions in the long-time 
limit. We have found two stable steady-state solutions accessible by time 
propagation. The first steady-state solution has densities $n_1=1.06$, 
$n_2=1.09$, the second one has $n_1=0.11$, $n_2=0.13$. The spectral functions 
corresponding to these solutions (see Fig.~\ref{fig:td_2b_dot}) are localized 
around the Fermi-level of the left or right lead respectively.
The inclusion of the long range interaction destroys the states where the first
peak is localized on the right lead energy band and the second peak is 
localized to the left lead energy band. Because the magnitude of the 
interaction felt by the electron on the site is now higher the density on 
the sites is decreased and the solution corresponding to the highest
density is at half-filling. Also in this case we are able to switch between
the two steady-states. The currents corresponding to these two
solutions for the density have almost the same magnitude.

In Fig.~\ref{fig:td_2b_dot} we show the time-dependent densities
and currents for the 2B approximation. Again within the correlated approximations
we find only one solution for the density and current. In the lower
panels of  Fig.~\ref{fig:td_2b_dot} we show the spectral functions
for the 2B and $GW$ approximations compared to the spectral functions of the HF approximation.
The 2B and $GW$  spectral functions are qualitatively quite different from 
 those of the HF approximation. Instead of the two peak structure of HF approximation, with 2B and $GW$ approximations we 
have one very broad peak with much lower maximum.

\begin{figure}[t]
	\centering
       \includegraphics[angle=0,width=\linewidth]{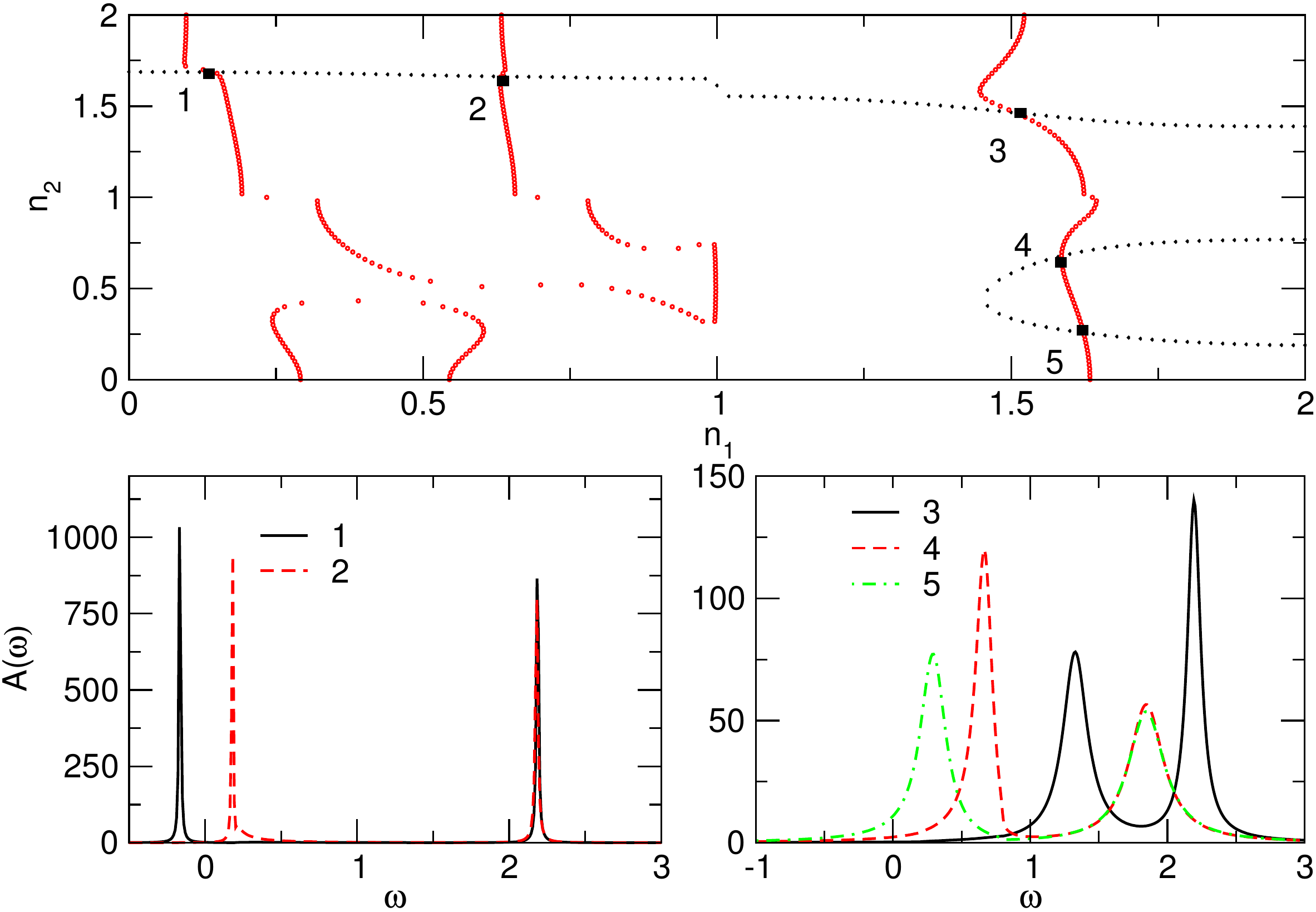}
       \caption[]{
         Upper panel: Graphical solution of the integral in Eq.~(\ref{selfcons_dens_dft}).
         Lower panel: Spectral functions for the BALDA with Hubbard interactions
         corresponding to the five different steady-state solutions for 
         the density.}
      \label{fig:n2-n1-balda}
\end{figure}  

\begin{table}
\begin{tabular}{|c|c|c|c|c|c|}
\hline
FP & $n_1$ & $n_2$ & FP & $n_1$ & $n_2$ \\ \hline
1 & 0.147 & 1.685 & 4 & 1.585 & 0.674 \\
2 & 0.632 & 1.658 & 5 & 1.624 & 0.250 \\
3 & 1.506 & 1.466 &  &  & \\ \hline
\end{tabular}
\caption{Fixed point (FP) solutions of Eq.~(\ref{selfcons_dens_dft}) for the 
steady-state densities of two interacting Hubbard sites connected to two 
biased, non-interacting leads in the BALDA approximation (see upper panel of 
Fig.~\ref{fig:n2-n1-balda}). 
The parameters are: $V_{\rm link}$ = 0.4, $W_{\textsc{L}}$ = 2.2,  
$W_{\textsc{R}}$ = -1.2, $U$ = 2.0, $V_{1,2}$ = 0.4, 
$\varepsilon_{\alpha} = \varepsilon_F = 0$, and  
$\varepsilon^{C}_{1} = -0.04$, $\varepsilon^{C}_{2}= 0.2$.  }
\label{table_ABALDA_fix}
\end{table}

We also studied the possibility of multiple steady-states for the 
same model within the BALDA. Using the same parameters as 
above,  the BALDA has multiple solutions. 
However, at least one fixed point 
has a density on one of the dots very close to unity, exactly 
where the BALDA potential is discontinuous. 
For a single interacting dot, 
this discontinuity has been shown to be closely related to the Coulomb 
blockade phenomenon.\cite{KurthStefanucciKhosraviVerdozziGross:10}
For the purposes of the present work we avoid the regime 
of integer occupancy in an ABALDA treatment by changing the on-site energies 
of the interacting sites in an asymmetric way such that 
$\varepsilon^{C}_1=-0.04$ and $\varepsilon^{C}_2=0.2$. With these 
modifications, the two coupled equations given by Eq.~(\ref{selfcons_dens_dft}), are solved 
simultaneously, yielding five fixed-points (see Fig.~\ref{fig:n2-n1-balda} 
and Table \ref{table_ABALDA_fix}). Among these five fixed points, FP 1, FP 3, 
and FP 5 are stable, the other two unstable. 

The spectral functions corresponding to FP 3 and FP 5 have two well separated 
smooth peaks, while the one corresponding to FP 1 has two sharp peaks, the 
first one located at $\w_1=-0.168$ outside the energy range of the left lead, 
the second one at $\w_2=2.16$ outside the energy range of the right lead. 

\begin{figure}[t]
	\centering
        \includegraphics[width=0.9\linewidth]{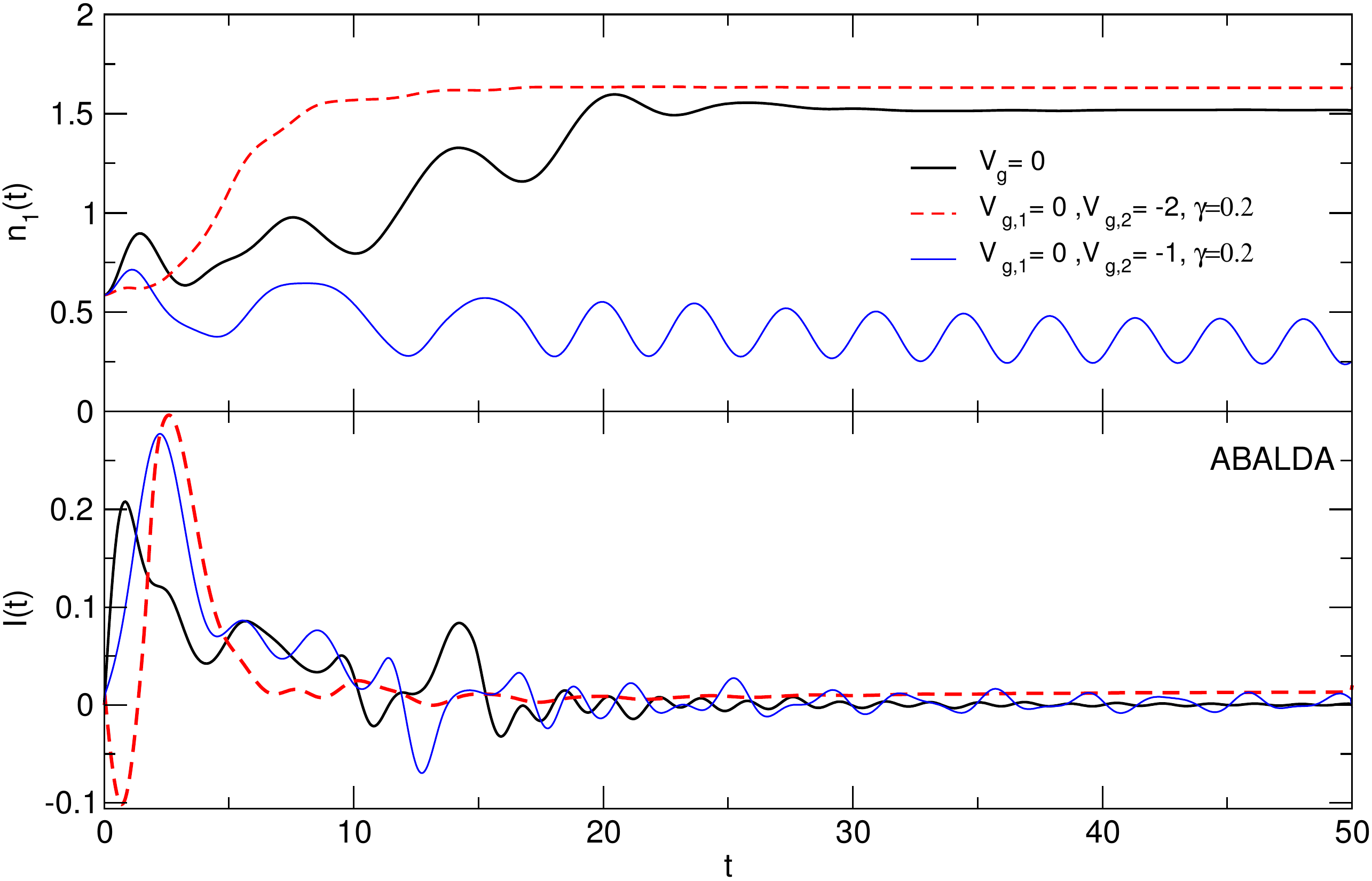}
        \caption[]{Time-dependent density and current for the ABALDA with 
                   different applied gates. }
        \label{fig:td_2b_dot2}
\end{figure}

Again, the stable solutions are accessible by time propagation. Upon 
application of a sudden bias in the leads at $t=0$, the system approaches the 
third solution if no external gate voltage is turned on.
On the other hand, if a gate voltage of the form (\ref{eq:gate}) is applied 
only to the second site, with amplitude $V_{g,2}=-2.0$ and $\g=0.2$, the 
system attains a steady-state with a density corresponding to FP 5.
As before switching between these two steady-state density is possible by 
changing the sign of the applied gate. A similar gate voltage applied only to 
the second site and smaller amplitude ($V_g=-1.0$) leads to an oscillatory 
time-dependent density, whose average total density is close to the one of 
FP 1. 

The frequency of the time-dependent density is $\w=2.24$ which is close to the 
energy difference ($\w=2.34$) between the peaks of spectral function. Hence, 
these oscillations are due to the existence of bound-states in
the biased interacting system. 
One possible way to explain the role of bound-states
in the biased KS Hamiltonian is that in the long-time 
limit the KS potentials are time dependent (with the bound state eigenenergy differences as
prominent frequencies) leading again to time-dependent
currents and densities (by virtue of Floquet's theorem). This
is evidently achieved in adiabatic approximations with the XC potential 
depending only on  the local  density such as the ABALDA and HF approximation.

\section{Conclusions}
\label{conclus}
In this paper we have investigated by means of real-time propagation within MBPT
and TDDFT, the existence of multiple steady states in single and double interacting 
quantum dot systems connected to semi-infinite leads. Within the framework of MBPT one 
can solve the steady-state MBPT equation without necessarily going through the whole 
time propagation. This can be done, by an iterative procedure like in Ref.~\onlinecite{ThygesenRubio:08}, for
example. In this case, starting from different initial guesses for the Green function,
bistability would manifest itself in the convergence to more than one self-consistent solution.
The advantage of the KB equations over the steady-state MBPT equations is that
during the transient regime the Green function explore a finite
portion of the domain of all possible Green functions. Thus
a single time propagation is similar to iteratively solving the steady-state
MBPT equations for a large number of initial guesses.
 
 In order to find the parameter region for bistability
we first solved the self-consistent steady-state equations within
the HF and BALDA approximation and determined the regime for which multiple solutions occur. We show
that only the stable solutions are accessible by time propagation. Moreover, we find that by
superimposing an exponentially decaying gate voltage pulse to the external bias, it is possible to
reach the various stable solutions and also to switch between them.
For the same parameters and driving fields, we then included dynamical 
XC effects by solving the Kadanoff-Baym equations with MBPT self-energies 
at the 2B and $GW$ level of approximation.
In all studied cases where adiabatic DFT and HF theory predict bistability 
{\em dynamical XC effects destroy the phenomenon}. 
Here we emphasize that we have performed 2B and $GW$ calculations for many 
more parameter sets than those for which we have shown results in the present 
work. We have found no indication for the existence 
of multiple steady states for any of these sets. 
However, due to the vastness of the parameter space, we cannot rule 
out completely the possibility of multiple steady states when dynamical 
XC effects are included.

 We wish to point out that even though ABALDA already contains correlations it is based on two approximations: 
the local and the adiabatic one. For any non-local but adiabatic 
approximation to the TDDFT functionals one could still derive a 
self-consistency condition for the steady-state density in the form of 
coupled, nonlinear equations. Because of this nonlinearity 
multiple solutions, i.e., multiple steady states, can be possible. 
Therefore our results suggest that it is the adiabatic approximation which 
permits bistability while we expect that the inclusion of memory effects suppresses it. 
We also conclude that bistable regimes induced by the 
electron-electron interaction only, are unlikely to be found in 
Hubbard or extended Hubbard model nanojunctions, and that 
other degrees of freedom, like molecular vibrations or nuclear 
coordinates, must be taken into account.

\begin{acknowledgments}
Part of the calculations were performed at the CSC -- IT Center for Science 
Ltd administered by the Ministry of Education, Science and Culture, Finland.
A.S. acknowledges funding by the Academy of
Finland under grant No. 140327/2010. 
S.K. acknowledges funding by the "Grupos Consolidados UPV/EHU del 
Gobierno Vasco" (IT-319-07).
R.v.L. acknowledges the Academy of Finland for research funding under
grant No. 127739.
We acknowledge the support from the 
European Theoretical Spectroscopy Facility. 
\end{acknowledgments}


\end{document}